\documentclass[preprint,12pt]{elsarticle}

\usepackage{url}
\usepackage{amssymb}
\usepackage{amsmath}
\usepackage{booktabs}
\usepackage{subcaption}
\usepackage[usenames]{color}
\newcommand{\teob}{\texttt{TEOBResum}~}
\newcommand{\seob}{\texttt{SEOBNR}~}
\newcommand{\phenom}{\texttt{IMRPhenom}~}

\journal{}

\begin{document}
\begin{frontmatter}

\title{Deep Residual Error and Bag-of-Tricks Learning for Gravitational Wave Surrogate Modeling}

\affiliation[inst1]{organization={Department of Informatics, Aristotle University of Thessaloniki},
            city={Thessaloniki},
            country={Greece}}
  \affiliation[inst2]{organization={Department of Physics, Aristotle University of Thessaloniki},
            city={Thessaloniki},
            country={Greece}}
            
 \affiliation[inst3]{organization={GSI Helmholtz Center for Heavy Ion Research},
            city={64291 Darmstadt},
            country={Germany}}

\author[inst1]{Styliani-Christina Fragkouli}
\ead{sfragkoul@certh.gr}

\author[inst1]{Paraskevi Nousi}
\ead{paranous@csd.auth.gr}

\author[inst1]{Nikolaos Passalis}
\ead{passalis@csd.auth.gr}

\author[inst3,inst2]{Panagiotis Iosif}
 \ead{piosif@auth.gr}

\author[inst2]{Nikolaos Stergioulas}
\ead{niksterg@auth.gr}

\author[inst1]{Anastasios Tefas}
\ead{tefas@csd.auth.gr}

\begin{abstract}

Deep learning methods have been employed in gravitational-wave astronomy to accelerate the construction of surrogate waveforms for the inspiral of spin-aligned black hole binaries, among other applications. We face the challenge of modeling the residual error of an artificial neural network that models the coefficients of the surrogate waveform expansion (especially those of the phase of the waveform) which we demonstrate has sufficient structure to be learnable by a second network. Adding this second network, we were able to reduce the maximum mismatch for waveforms in a validation set by 13.4 times. We also explored several other ideas for improving the accuracy of the surrogate model, such as the exploitation of similarities between waveforms, the augmentation of the training set, the dissection of the input space, using dedicated networks per output coefficient and output augmentation. In several cases, small improvements can be observed, but the most significant improvement still comes from the addition of a second network that models the residual error. Since the residual error for more general surrogate waveform models (when e.g., eccentricity is included) may also have a specific structure, one can expect our method to be applicable to cases where the gain in accuracy could lead to significant gains in computational time. 
\end{abstract}


\end{frontmatter}
\section{Introduction}
\label{sec:introduction}

\noindent \emph{Observations of GW and the global network of detectors.} Since the first detection of gravitational waves (GW) from a system of binary black holes (BBH) in 2015 \cite{Abbott:2016blz}, GW detections have become more and more frequent, approaching gradually the status of routine observations. From that date onwards, GW detections have become more and more frequent, approaching gradually the status of routine observations. The initial observing run (O1) of the two Advanced LIGO \cite{TheLIGOScientific:2014jea} laser interferometers based in Hanford and Livingston, resulted in 3 GW events. During the second run (O2) the total number of registered detections increased to 11 (first Gravitational-Wave Transient Catalog, GWTC-1, \cite{LIGOScientific:2018mvr}), with the Advanced Virgo detector \cite{acernese2014advanced} joining in towards the end of that observational period. Virgo's addition to the twin LIGO detectors was paramount, as it coincided with the first detection of a binary neutron star (BNS) coalescence, GW170817 \cite{LVK_2017PhRvL.119p1101A}, accompanied with extensive observations in the electromagnetic (EM) spectrum \cite{Abbott_2017, GW170817_EM_counterpart}. The first part of the third observing run (O3a) updated the number of events to 50 (GWTC-2, \cite{Abbott:2020niy}), while the latest catalog (GWTC-3, \cite{GWTC3_LVK_2021}) contains 90 GW events.

Continued improvements in the detectors' sensitivity \cite{abbott2020prospects} are expected to further increase the number of GW observations. A fourth GW observatory, KAGRA \cite{akutsu2019kagra, KAGRA_2021}, joined the global network of GW detectors at the end of the O3 run. KAGRA will participate in the fourth observing run (O4) which is planned to begin in May 2023, while also improving its sensitivity during O4. Additional observatories are important so that the sky localization of GW sources is more accurate and their properties are determined with higher precision, thus providing crucial information for potential EM follow-up observations of GW events. To that end, the construction and operation of a fifth interferometer, LIGO-India \cite{LIGO_India_2022} will also significantly improve both the network sensitivity and the sky localization. Furthermore, third generation ground-based detectors such as the Einstein Telescope \cite{Punturo_etal_2010_ET, Maggiore_etal_2020_ET} and Cosmic Explorer \cite{Reitze_etal_2019_CE, Evans_etal_2021_CE} are actively being developed and they are anticipated to vastly improve our knowledge of astrophysical processes in the Universe  \cite{NextGen_detectors_2017, GWIC_3G_reports_intro_2021, GWIC_3G_reports_science_2021}.

\vspace{0.5cm}

\noindent \emph{GW modeling.}
The breakthroughs in GW astronomy briefly described above, were made possible thanks to collaborative efforts on multiple fronts. The rigorous developments in gravitational waveform modeling of a variety of compact binary coalescences (CBC) was undoubtedly a crucial point. A prerequisite to interpret GW signals and unveil the properties of their source is to solve the two-body problem in General Relativity (GR). To that end, one needs to solve the Einstein equations and this can be done either  via analytical methods (truncated at some order in a post-Newtonian (PN) expansion, see e.g. \cite{Blanchet_2014}) or via numerical solutions of the full Einstein equations (with some assumptions and up to a specified tolerance). The analytical approach is fast, but it is valid only for large separations of the binary and small (compared to the speed of light) orbital velocities, i.e. it breaks down after the late inspiral phase. The subsequent merger and ringdown phase of the coalescence, i.e. the strong field and relativistic velocities regime, can be captured accurately \textit{only} with numerical relativity (NR). However, the computational cost of such NR simulations (typically performed on supercomputers) is tremendous, ranging from tens of thousands to millions of corehours \cite{PhysRevLett.96.111101, PhysRevLett.95.121101, PhysRevLett.96.111102, Boyle_etal_2019_SXS_catalog, Dietrich_etal_2018_CoRe_database}.

Efforts to combine the two aforementioned approaches in order to produce accurate descriptions of the \textit{entire} coalescence, including the full inspiral, merger and ringdown, have resulted in different families of waveform models. The main ones are the effective-one-body (EOB) and the phenomenological families. The former has two prominent members, the \seob models \cite{Buonanno_etal_2007, 2023arXiv230318046R}, and the \teob models \cite{Damour_Nagar_2014, Gamba_etal_2021_arXiv211103675G}, while the latter comprises of the \phenom models \cite{Ajith_etal_2008, pratten2021IMRPhenomXPHM}.

The EOB approach \cite{Buonanno_Damour_1999} generalizes to GR the Newtonian result dictating that the relative motion of a two-body system is equivalent to the motion of a particle of mass $\mu = m_1 m_2/(m_1 + m_2)$ (where $m_1$ and $m_2$ are the masses of the binary components) in the two-body potential $V(r)$ \footnote{Specifically, the GR two-body problem becomes the problem of describing the evolution of a test mass orbiting around a deformed Kerr metric.}. In the EOB context, the PN inspiral information is resummed and calibrated to NR data and the merger-ringdown part is obtained from a fit to NR data. The two EOB subfamilies, \seob and \teob,  differ mainly regarding the choices in the resummation process and in the amount of PN and NR information employed.

In the phenomenological approach, the binary coalescence is typically split into 3 regions, where piecewise closed form expressions are used to represent the waveforms. Breaking up the total frequency region into: (i) a low frequency inspiral regime, where the waveform can be described by adding additional terms to a PN expansion, (ii) a high frequency regime, where the waveform is dominated by a quasi-normal ringdown that can be described by perturbation theory and (iii) an intermediate regime, which captures the complex physics of the merger and demands insights from NR, facilitates the finding of appropriate analytical functions. These expressions are fitted to NR waveform data for a set of constructed \textit{hybrid} waveforms that in essence ``glue'' together EOB waveforms describing the inspiral and NR waveforms describing the last orbits, merger and ringdown of the binary.

The latest implementations of the EOB and phenomenological families of models include effects from the spin-induced precession of the binary orbit and contributions from both the dominant and subdominant multipole moments of the emitted gravitational radiation. 
The current state-of-the-art models are \texttt{SEOBNRv5PHM} \cite{2023arXiv230318046R} for the \seob family, \texttt{IMRPhenomXPHM} \cite{pratten2021IMRPhenomXPHM} and \texttt{IMRPhenomTPHM} \cite{Estelles_etal_2021_arXiv210505872E} for the \phenom family and \texttt{TEOBResumS-GIOTTO} \cite{Gamba_etal_2021_arXiv211103675G} for the \teob family.
Furthermore, tidal effects can be incorporated \cite{Dietrich_etal_2017, dietrich2019tidal, dietrich2019improved} in the \seob \cite{Matas_etal_2020} and \phenom \cite{Thompson_etal_2020} models as appropriate, while in the \teob family tidal and spin effects are merged together into a single EOB framework and therefore no extension is needed as in the other two families.

Progress towards the above cutting-edge waveform models has followed a course of stepping stones along which extra features (i.e. inclusion of spins, higher modes, precession) and different optimization and acceleration techniques were implemented gradually. Early efforts in the EOB front (e.g. \cite{Damour_2001, Damour_etal_2008} resulted in the first versions of the \seob models which evolved from describing non-spinning binaries to spinning, precessing systems. \cite{pan2011inspiral, Taracchini_etal_2012, Pan_etal_2014, Taracchini_etal_2014}.
Improved EOB waveform models soon followed \cite{bohe2017improved, Babak_etal_2017, cotesta2018, ossokine2020, Cotesta_etal_2020} leading up to the 5th and most sophisticated implementation of \seob models \cite{2023arXiv230318046R, 2023arXiv230318039P, 2023arXiv230318026V, 2023arXiv230318143K}. Similarly, groundwork for the \phenom family of waveform models \cite{Schmidt_etal_2012, Hannam_etal_2014} developed to include higher modes \cite{mehta2017, london2018, mehta2019, khan2020}. Further improvements in accuracy \cite{pratten2020domharm, garciaquiros2020, Estelles_etal_2021, GarciaQuiros_2021} led to the latest generation of \phenom models \cite{pratten2021IMRPhenomXPHM, Estelles_etal_2022_PhysRevD.105.084039} becoming a standard tool in GW parameter estimation. The \teob family of models followed a similar course of incremental development \cite{Bernuzzi_etal_2015, nagar2018, nagar2020, nagar2020teob, Akcay_etal_2021}.
In addition, the expansion of the above waveform family trees from quasi circular to eccentric binaries is diligently pursued \cite{Setyawati_Ohme_2021, Chiaramello_Nagar_2020, Nagar_etal_2021_general, Khalil_etal_2021, RamosBuades_etal_2021}.

On one hand, the inclusion of extra physical characteristics as outlined above, allows for more realistic and complex waveforms, which is a prerequisite to perform highly accurate parameter estimation of GW sources. Such state-of-the-art models have played a significant role in the analysis of GW from: (i) asymmetric mass binaries \cite{LIGOScientific:2020stg, Abbott:2020khf, Colleoni_etal_2021}, (ii) systems whose binary components include one or two neutron stars \cite{LVK_2020_GW190425, LVK_2021_NSBH}, (iii) GW190521 \cite{LVK_2020_GW190521} the most massive BBH merger detected to date \cite{Estelles_etal_2022_GW190521, Nitz_Capano_2021} and (iv) independent re-analysis of the public LIGO-Virgo-KAGRA (LVK) data, e.g. \cite{Nitz_etal_2021_4OGC}.

On the other hand, the increased complexity decreases the waveforms' computational efficiency (e.g. compared to the simpler PN approximants). This effect is more pronounced for EOB waveforms, as phenomenological waveforms are faster by construction. The computational cost of EOB models is burdened by the need to tackle the orbital dynamics through solving a complex system of ordinary differential equations. A solution to this problem has been supplied by surrogate modeling\footnote{Complementary to the surrogate solution, recently, the EOB description of the binary dynamics has been bolstered with a technique called the post-adiabatic approximation which promises to speed up the waveform generation to levels computationally competitive with current phenomenological and surrogate models \cite{Nagar_Rettegno_2019_PA, Gamba_etal_2021, Riemenschneider_etal_2021, Mihaylov_etal_2021}.} \cite{field2014fast, Tiglio_Villanueva_2021_review_arXiv210111608T}. By fitting interpolated decomposed waveform data pieces over the binary parameter space, surrogate models can significantly accelerate either NR (e.g. \cite{Blackman_etal_2015, blackman2017surrogate, Blackman_etal_2017b, varma2019, varma2019precess, Islam_etal_2021}) or EOB waveforms (e.g. \cite{field2014fast, Purrer_2016, Lackey_etal_2017, Lackey_etal_2019, Yun_etal_2021}), while maintaining high accuracy within their parameter space of validity.

\vspace{0.5cm}

\noindent \emph{Motivation and Contribution.}
In this study, we focus on the SEOBNRv4 model, which has a 3-dimensional parameter space $\boldsymbol{\lambda}$; the mass ratio $q$ between the two BHs and their spins $\chi_1$ and $\chi_2$, assuming that they are aligned with the orbital angular momentum. Surrogate models have been shown to be fast and reliable approximations of waveforms such as SEOBNRv4, within a specified tolerance error. A surrogate model for this family of waveforms is presented in \cite{khan2021gravitational}. 

As discussed in \cite{khan2021gravitational} there are challenges in modeling GW signals and there is a need to reduce computational costs to increase the efficiency of analyzing GW events. While there have been significant advancements in GW signal modeling, incorporating more complex features increases computational costs, limiting their use. To overcome this challenge, custom-made optimizations have been developed \cite{Nagar_Rettegno_2019_PA}, but they require expert knowledge and may not provide general optimizations. Alternatively, data-driven methods such as surrogate modeling can be employed, which provide accurate approximations of computationally expensive waveform models \cite{field2014fast, nousi2021autoencoderdriven}. Several machine learning techniques are available for interpolating or fitting the projection coefficients of a reduced basis representation of time-domain waveforms, and the most appropriate method depends on the required accuracy and dimensionality. For low-dimensional parameter spaces, interpolation is a viable option. However, as dimensionality increases, interpolation becomes challenging due to the substantial number of data points usually required. ANNs are proposed to estimate these coefficients since this approach allows for efficient execution on either a CPU or GPU.

In reproducing the results of \cite{khan2021gravitational}, we noticed that the residual errors after training the neural networks had structure with respect to the input parameters and hypothesized that a second neural network could learn to model these errors. Our contribution can be summarized as follows:
\begin{itemize}
    \item We design and train a neural network to model the residual errors of the surrogate model network.
    \item Drawing inspiration from the physical aspect of the task at hand, we introduce various tricks to reduce the learning errors.
    \item We make use of the residual errors of the training of the ANN model and achieve a maximum mismatch between SEOBNRv4 waveforms and waveforms generated by our surrogate model that is more than one order of magnitude smaller compared to the baseline method.
\end{itemize}
To the best of our knowledge, this is the first work to model the residual errors of a neural network based gravitational wave surrogate model. The code for this work is available at \url{https://github.com/sfragkoul/residual-gw-surrogate-modelling}.

The rest of this paper is arranged as follows: in Section~\ref{related} related works are detailed and compared to this work, and in Section~\ref{proposed} surrogate modeling is briefly overviewed, the ground truth used during our experiments is described and
the baseline network is
presented.
In Section~\ref{sec: alternations} we discuss the improvements noticed by adding a supplementary residual error learning network, i.e. a network predicting the errors of the training network. In Section~\ref{sec: exploration} we summarize different approaches investigated to manipulate the input and output space of the networks, aiming to obtain a better mismatch.
Section~\ref{sec: results} the presentation of how the variants of the ANN surrogate model, as described in the previous sections, perform takes place, and in Section~\ref{conclusions} we summarize our findings. Finally, in the Appendix we show additional results for the distribution of the mismatch values for different values of the greedy tolerance.

\section{Related Work}
\label{related}

Interest in harnessing machine learning techniques for the analysis of GW data has been on the rise lately (see \cite{cuoco2020review} for a review). For example, the interpolation (i.e. the third step described above) can be very costly at high dimensions, in which case probabilistic methods could be employed, see e.g. \cite{2021arXiv211008901B}. In \cite{chua2019reduced} the complex waveforms are separated into the real and imaginary segments and two separate surrogate models are constructed, with artificial neural networks (ANNs)  implemented within a 4-dimensional input space to fit the coefficients from the reduced basis, omitting the final step of the empirical interpolation.

In \cite{setyawati2020regression}, various interpolation methods were investigated, concluding that machine-learning based methods may perform better as the complexity of the surrogate modeling problem increases. In another important development, using genetic programming and symbolic regression techniques, surrogate models based on numerical simulations (i.e. essentially in an ab-initio approach) were used to obtain closed-form expressions for the GW emission of BBH collision, modeling the entire coalescence \textit{at once}, i.e. without the need to distinguish between the different regimes of inspiral, merger and ringdown \cite{Tiglio_Villanueva_2021NatSR}.

In \cite{khan2021gravitational} a time-domain surrogate model of the spin-aligned BBH waveform model SEOBNRv4 \cite{bohe2017improved} was built by utilizing ANNs instead of interpolating during the last step, and also by dividing the problem to the signal parts of the phase and amplitude, which resulted in the creation of two corresponding surrogate models. Furthermore, another recent application of Deep Learning (DL) was implemented  \cite{nousi2021autoencoderdriven}, where ANNs were used (specifically autoencoders), for the examination of latent structures in the coefficients from the empirical interpolation.

In \cite{Barsotti_2022} the authors present the prospect for predicting gravitational waveforms from compact binaries based on automated machine learning (AutoML). The study focuses on the analysis of GW emitted when two spinless black holes collide in an initial quasi-circular orbit. Their findings suggest that AutoML has the potential to serve as a framework for regression in the field of surrogates for gravitational waveforms. Specifically, this research is conducted in the context of surrogates derived from NR simulations using the reduced basis and the empirical interpolation. Their results indicate that AutoML is capable of generating surrogates that closely resemble the actual NR simulations. In \cite{PhysRevD.106.104029}, the use of ANNs for surrogate modeling is investigated and it is shown that they can be used to build computationally efficient and accurate models for multipolar waveform models of precessing BBHs. The method is applied on the \texttt{SEOBNRv4PHM} waveform model and the generated surrogate model is faster than the original model by 2 orders of magnitude on a CPU. With batched operations and the acceleration of a GPU, this speedup is further increased.

The main advantage of using an ANN-surrogate model in our study is its ability to capture complex nonlinear relationships between the input parameters and the output waveform signals, which can be challenging to model using traditional analytic methods. This approach has been shown to be effective in previous studies for predicting gravitational waveforms \cite{khan2021gravitational}, and we evaluated the performance of our proposed ANN-surrogate model against a baseline model, which was also based on an ANN architecture. We chose this approach for comparison as it allowed us to assess the effectiveness of our proposed method in improving the accuracy of the predictions relative to a baseline.

Unlike the aforementioned works \cite{khan2021gravitational, nousi2021autoencoderdriven}, though, in this study we take into account modeling errors generated by the DL models by learning the residual errors. Furthermore,  we draw inspiration from the physical aspect of the task at hand as well as the learning objective, and introduce various tricks to reduce the learning errors. Notably, we make use of the residual errors of the training of the ANN model in order to achieve significantly better results.

\section{Background}
\label{proposed}

In the following subsections we provide the theoretical background of surrogate modeling for gravitational waves. Furthermore, we discuss the specifics of constructing an effective surrogate model for the SEOBNRv4 waveform family, following the steps described in \cite{khan2021gravitational}.

\subsection{Construction of a surrogate model}
\label{sec:surrogate}

We use the notation $h(t;\boldsymbol{\lambda}) = h_+(t;\boldsymbol{\lambda}) - ih_{\times}(t;\boldsymbol{\lambda})$ to denote a complex gravitational wave strain, where $h_+$ and $h_{\times}$ are the two independent polarizations \cite{maggiore}, $t$ denotes time and the elements of $\boldsymbol{\lambda}$ are the intrinsic parameters. In general, for inspiraling black holes in general relativity on non-eccentric orbits, these parameters are 7-dimensional, consisting of the mass ratio and two spins with arbitrary orientation. Under restrictions, these parameters can be simplified to only include the mass ratio and two spins aligned with the orbital plane. We study this case, based on the SEOBNRv4 model specifically \cite{bohe2017improved}.

Surrogate modeling aims to approximate the given signals using a reduced model, denoted as $h_{s}(t;\boldsymbol{\lambda})$, such that the approximation given by the surrogate model, $h_{s}(t;\boldsymbol{\lambda})$ accurately reconstructs the actual waveform $h(t;\boldsymbol{\lambda})$ within a preset threshold of error. When considering only the dominant, quadrupole ($l=m=2$) mode \cite{maggiore}, the target becomes $h_{s}(t;\boldsymbol{\lambda}) \approx h_{2,2}(t;\boldsymbol{\lambda})$ where $l,m$ are the spherical harmonics.
In surrogate modeling, the first step is to prepare a large {\it training set} of waveforms, as previously mentioned. Implementing the SEOBNRv4 model \cite{bohe2017improved}, whose input space is three-dimensional, each waveform is parameterized by the mass ratio $q\equiv{m_1}/{m_2}\geq 1$ and the dimensionless spins $\chi_1, \chi_2$ of the two black holes. Thus, a training set of $N$ waveforms $\{ h_i(t;\boldsymbol{\lambda}_i) \}_{i=1}^{N}$  is created\footnote{The \texttt{PyCBC} package \cite{pycbc} was used to generate the waveforms, internally calling methods from \texttt{LALSuite} \cite{lalsuite}.}, where 
${\boldsymbol{\lambda}}_i = (q, \chi_1, \chi_2)_i$. The mass ratio is limited to a predetermined interval, e.g. as $1\leq q \leq 8$, inside which the surrogate model is accurate by design. For the two spins, their values can be in the range $-0.99\leq \chi_{1,2} \leq 0.99$.

A {\it reduced basis} is built from the training set, using either a greedy algorithm \cite{field2014fast} or algebraic approaches, like Singular Value Decomposition (SVD) \cite{blackman2017surrogate}.
The greedy algorithm is an iterative process, which chooses $n < N$ waveforms (and, by extension, their corresponding ${\{\boldsymbol\lambda}_j\}_{j=1}^{n}$ values, the greedy points), which, after orthonormalization, constitute the reduced basis $\{e_j\}_{j=1}^{n}$. Each $\boldsymbol{\lambda}_i$ waveform in the training set is then represented as a linear combination 
\begin{equation}
    h(t;\boldsymbol{\lambda}_i) \approx \sum_{j=1}^{n} c_j(\boldsymbol{\lambda}_i) e_j(t),
\end{equation}
within a preset error tolerance, where  $\{c_j(\boldsymbol{\lambda}_i)\}_{j=1}^{n} = \left\langle h(t ; \boldsymbol{\lambda}_i), e_{j}(t)\right\rangle$ are the orthogonal projection coefficients. To ensure that our coefficients remain smooth with respect to the parameters, we adopted measures to mitigate the impact of noisy data. Specifically, we employed a higher sampling rate and a larger number of training samples, which enabled us to obtain a smooth and dense input parameter space.  Furthermore, our data are generated by highly-accurate semi-analytic waveform models, so they do not suffer from experimental data noise -- any numerical data noise is orders of magnitude smaller than the mismatch we are aiming for. Our approach is in line with the recommendations put forth by \cite{field2014fast}.

Next, a new EIM basis $B_k(t)$ is obtained, such that a waveform  $h(t;\boldsymbol{\lambda}_j)$ (where $\boldsymbol{\lambda}_j$ is one of the greedy points of the ROM basis) can be represented as the linear combination
\begin{equation}
    h\left(t ; \boldsymbol{\lambda}_j\right)=\sum_{k=1}^{n}  \alpha_k(\boldsymbol{\lambda}_j) B_{k}(t),
    \label{eq:EIM-reconstruction}
\end{equation}
where the coefficients $\alpha_k(\boldsymbol{\lambda}_j)$ coincide with the waveform at particular times, $\{ T_k \}_{k=1}^{n}$, the  \emph{empirical time nodes}, i.e. $\alpha_k(\boldsymbol{\lambda}_j) =h\left(T_{k} ; \boldsymbol{\lambda}_j \right)$.

For any other waveform $h\left(t ; \boldsymbol{\lambda}_i\right)$ in the training set, the coefficients of the EIM representation are simply  $\alpha_k(\boldsymbol{\lambda}_i) =h\left(T_{k} ; \boldsymbol{\lambda}_i \right)$. Since this does not involve the basis $B_k(t)$, the coefficients are computed much faster (as known values of the waveform at particular times) than the projection coefficients in the ROM basis (which require the projection of the whole waveform).

A surrogate model is finally produced, by interpolating over the coefficient matrix $\alpha_k(\boldsymbol{\lambda}_i)$ of the training set to find the coefficients $\hat \alpha_k(\boldsymbol{\lambda})$ for an arbitrary $\boldsymbol{\lambda}$, such that
\begin{equation}
    h\left(t ; \boldsymbol{\lambda}\right)\approx \sum_{k=1}^{m} \hat \alpha_{k}(\boldsymbol{\lambda}) B_{k}(t).
    \label{eq:EIM}
\end{equation}
Depending on the dimensionality of $\boldsymbol{\lambda}$,  multi-dimensional interpolation is required, the computational cost of which increases dramatically with increasing number of parameters. {\it This particular part of the process can be accelerated using neural networks, as shown in} \cite{khan2021gravitational}.

In practice, we used a similar setup as in \cite{khan2021gravitational}, as our baseline model. Instead of working with the strain amplitudes $h_+$ and $h_\times$, we use the {\it amplitude} $A$ and {\it phase} $\phi$ of the complex waveform, defined through 
\begin{equation}
    h_+\left(t ; \boldsymbol{\lambda}\right) - h_{\times}\left(t ; \boldsymbol{\lambda}\right) = A\left(t ; \boldsymbol{\lambda}\right) e^{-i\phi\left(t ; \boldsymbol{\lambda}\right)},
\end{equation}
as this results in a more compact EIM basis.
The training set comprised  $N=2 \times 10^5$ waveforms, randomly sampled in the $1\leq q\leq 8, -0.99\leq \chi_{1,2}\leq 0.99$ parameter space. The waveforms in the training set were accurately aligned in amplitude and initial phase, the phase was unwrapped and the time series was truncated to correspond to a common starting time of $-20000M$, where we chose a total mass of $M=60M_\odot$. This ensured that all waveforms started with a minimum frequency no larger than 15 Hz. We kept $100M$ of post-peak ringdown data. The ROM and EIM bases were constructed using {\tt RomPy} \cite{field2014fast,rompy}.

A {\it validation set} of $3\times 10^4$ SEOBNRv4 waveforms (not included in the training set) was used to evaluate the accuracy of the reconstructed waveforms.

\subsection{Measuring the reconstruction error}

For two waveforms with parameters $\boldsymbol{\lambda}_1$ and $\boldsymbol{\lambda}_2$, one defines the inner product  \cite{flanagan}
\begin{equation}
    \langle h(\cdot; \boldsymbol{\lambda}_1), h(\cdot; \boldsymbol{\lambda}_2) \rangle = 4\Re\int_{f_{min}}^{f_{max}} \frac{\tilde h(f;\boldsymbol{\lambda}_1) \tilde h^{*}(f;\boldsymbol{\lambda}_2)}{S_{n}(f)} df,
\end{equation}
where $\tilde h(f;\boldsymbol{\lambda})$ is the Fourier transform of $h(t;\boldsymbol{\lambda})$, $S_{n}(f)$ denotes the noise power spectral density (PSD) of the GW detector\footnote{We used the PSD of the Advanced LIGO design sensitivity \cite{noise}. There is a minimal impact on the calculated mismatch, with respect to the choice of a flat PSD.} and the star notation denotes the complex conjugate. The inner product can be used to {\it normalise} the Fourier transform of a waveform as
\begin{equation}
    \hat{h}(f; \boldsymbol{\lambda}) = \frac{{\tilde h}(f; \boldsymbol{\lambda})}{\langle h(\cdot; \boldsymbol{\lambda}), h(\cdot; \boldsymbol{\lambda}) \rangle},
\end{equation}
Then, the {\it overlap} between two waveforms is defined as the inner 
product between normalised waveforms $\hat{h}(\cdot; \boldsymbol{\lambda}_1)$, $\hat{h}(\cdot; \boldsymbol{\lambda}_2)$, maximised over a relative time ($t_0$) and phase ($\phi_0$) shift between
the two waveforms:
\begin{equation}
    {\cal O}(\hat{h}(\cdot; \boldsymbol{\lambda}_1), \hat{h}(\cdot; \boldsymbol{\lambda}_2)) = \max_{t_0, \phi_0} \langle h(\cdot; \boldsymbol{\lambda}_1), h(\cdot; \boldsymbol{\lambda}_2) \rangle,
\end{equation}
and, finally the {\it mismatch} is given by
\begin{equation}
    \label{eq:mismatch}
    \mathcal{M}(\hat{h}(\cdot; \boldsymbol{\lambda}_1), \hat{h}(\cdot; \boldsymbol{\lambda}_2)) = 1 - {\cal O}(\hat{h}(\cdot; \boldsymbol{\lambda}_1), \hat{h}(\cdot; \boldsymbol{\lambda}_2)) .
\end{equation}
To measure the performance of the surrogate model, the mismatch between actual waveforms, given by the SEOBNRv4 model, and surrogate predictions is used.

\begin{table}
\caption{\label{tab:mismatches}
Number of coefficients ($n$) of the reduced EIM bases for the amplitude and phase for different values of the greedy tolerance. The last three columns show the mismatch $\mathcal{M}$ (maximum, median and $95^{\rm th}$ percentile values) of the waveforms in the validation set when reconstructed via Eq. (\ref{eq:EIM-reconstruction}).}
\centering
\resizebox{1 \textwidth}{!}{
\begin{tabular}{|c c c c c c|} 
 \hline
 Greedy & $n$  & $n$  &   & mismatch $\mathcal{M}$ & \\ [0.1ex] 
 Tolerance & (amplitude) & (phase) & (max) & (median) & ($95^{\rm th}$ percentile) \\[0.1ex] 
 \hline
 $10^{-6}$ & $8$ & $4$ & $8.44\times10^{-3}$ & $5.47\times10^{-4}$ & $1.81\times10^{-3}$\\ 
 $10^{-8}$ & $13$ & $4$ & $8.44\times10^{-3}$ & $5.45\times10^{-4}$ & $1.80\times10^{-3}$\\
 $ 10^{-10}$ & $18$ & $8$ & $4.95\times10^{-4}$ & $1.30\times10^{-5}$ & $8.22\times10^{-5}$ \\
 $10^{-12}$ & $41$ & $12$ & $2.07\times10^{-6}$ & $7.45\times10^{-8}$ & $2.83\times10^{-7}$ \\
 $10^{-14}$ & $84$ & $32$ & $1.34\times10^{-8}$  & $5.64\times10^{-10}$ & $3.95\times10^{-9}$\\ 
  $10^{-16}$ & $93$ & $48$ & $6.60\times10^{-9}$  & $4.59\times10^{-10}$ & $3.02\times10^{-9}$\\
 \hline
\end{tabular}}

\end{table}

\begin{figure}
     \centering
     \begin{subfigure}[ht]{0.55\textwidth}
         \centering
         \includegraphics[width=\textwidth]{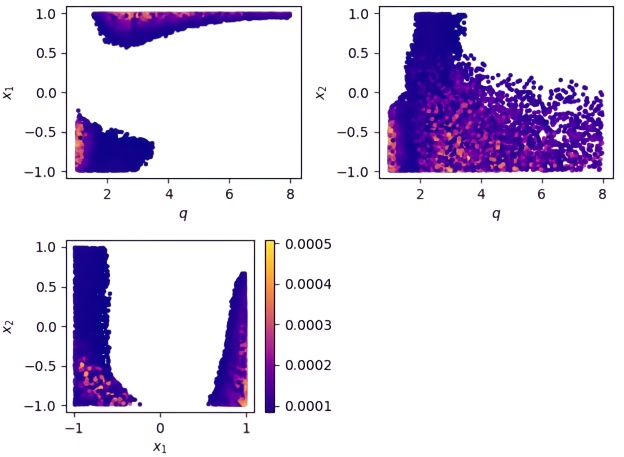}
         \label{train_gt}
     \end{subfigure}
     \hfill
     \begin{subfigure}[ht]{0.4\textwidth}
         \centering
         \includegraphics[width=\textwidth]{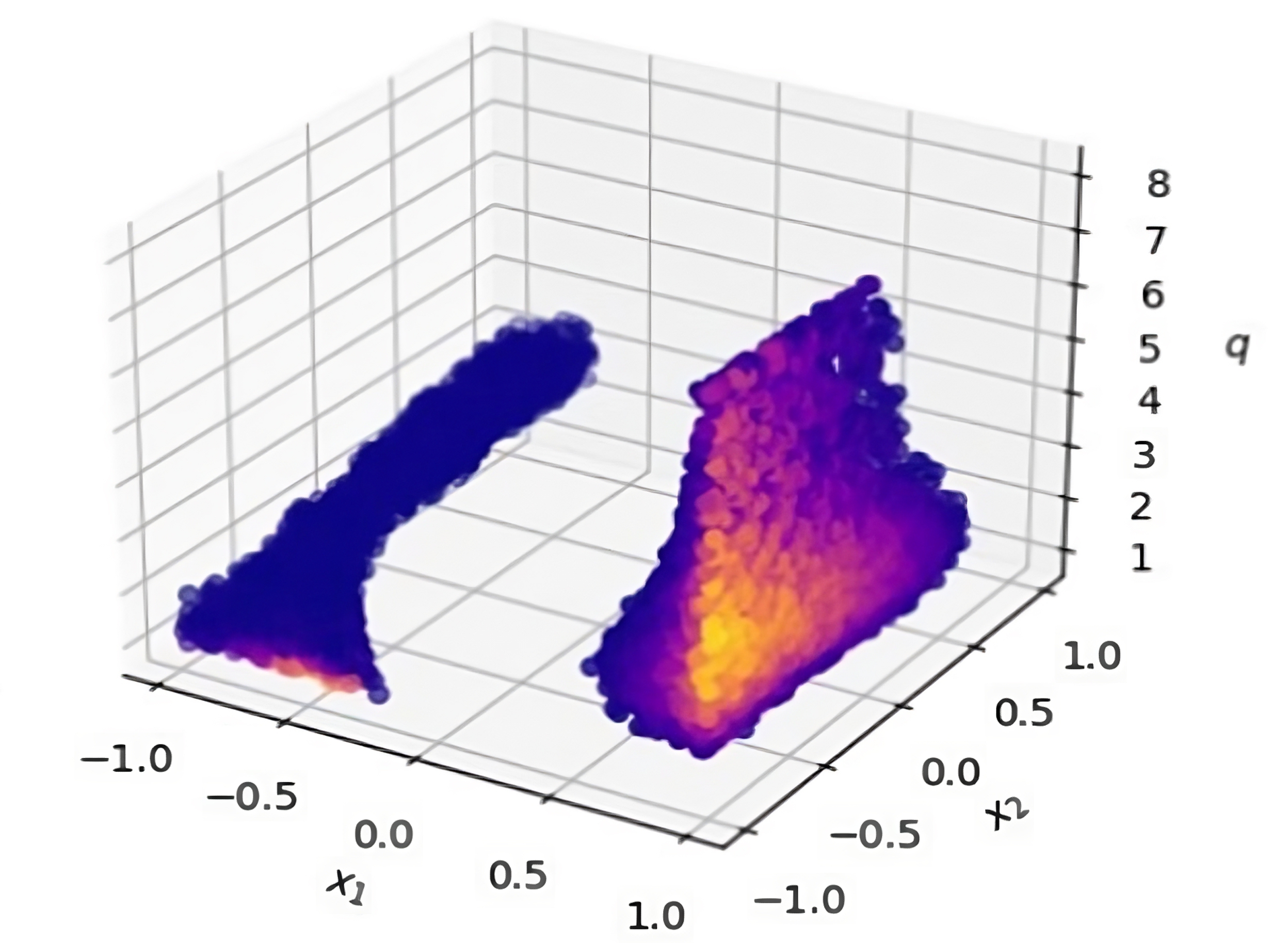}
         \label{train_gt_3d}
     \end{subfigure}
     \hfill
        \caption{Two-dimensional and three-dimensional distribution of mismatch values larger than $\sim 8\times 10^{-5}$ ($95^{th}$ percentile) for the waveforms in the training set, when reconstructed using the EIM reduced basis via Eq. (\ref{eq:EIM-reconstruction}), with a greedy tolerance of $10^{-10}$.}
        \label{mismatch-dist-train}
\end{figure}

\subsection{Performance of the reduced basis reconstruction and ground truth}
\label{sec:reduced_basis}

Table \ref{tab:mismatches} shows the number of coefficients of the reduced EIM basis for the amplitude and for the phase for different values of the greedy tolerance. The last three columns show the maximum, median and $95^{\rm th}$ percentile mismatch of
the waveforms in the validation set when reconstructed via Eq. (\ref{eq:EIM-reconstruction}). An initial greedy tolerance of $10^{-6}$ results in a basis with just 8 coefficients for the amplitude and 4 coefficients for the phase, with a median mismatch of the ANN surrogate model of $\simeq 5\times 10^{-4}$. Decreasing the greedy tolerance to $10^{-10}$ reduces the  median mismatch to $\simeq 10^{-5}$, requiring  18 coefficients for the amplitude and 8 coefficients for the phase. With even lower greedy tolerances, one can reach a median mismatch of only $\simeq 5\times 10^{-10}$ at the expense of a larger number of coefficients. The results in  Table \ref{tab:mismatches} are comparable to the corresponding results in Table I of \cite{khan2021gravitational}. 

For the case of a greedy tolerance of $10^{-10}$, the two-dimensional and three-dimensional distributions of mismatch values larger than $\sim 8\times 10^{-5}$ ($95^{th}$ percentile) for the waveforms in the training set, when reconstructed using the EIM reduced basis via Eq. (\ref{eq:EIM-reconstruction}), are shown in Fig. \ref{mismatch-dist-train}. The corresponding distributions for the validation set and for various values of the greedy tolerance are shown in the Appendix. The distributions are different at very small values of the greedy tolerance, when compared to the distributions for higher values of the greedy tolerance.

Next, selecting a greedy tolerance of $10^{-10}$, the $k=18$ coefficients 
$\alpha_k(\boldsymbol{\lambda}_i) =h\left(T_{k} ; \boldsymbol{\lambda}_i \right)$
for the amplitude and corresponding 8  coefficients for the phase (for each of the $N=2 \times 10^5$ waveforms with parameters $\boldsymbol{\lambda}_i$ in the training set) were used as the {\it ground truth} for the baseline ANN surrogate model discussed in the next subsection. In the remainder of this paper, we will denote the ground truth EIM coefficients of the $N$ training set waveforms as 
$\boldsymbol{y}_i \equiv \{ \alpha_k(\boldsymbol{\lambda}_i) \}_{k=1}^{n}$.

\subsection{ANN surrogate model: the baseline network}
\label{sec:baseline_impl}

To complete the surrogate model, an ANN was trained to interpolate the coefficients  $\alpha_k({\boldsymbol{\lambda}_i})$ of the training set to find the coefficients $\hat \alpha_k(\boldsymbol{\lambda})$ for an arbitrary $\boldsymbol{\lambda}$. We implemented the idea and baseline ANN model which followed the architecture from \cite{khan2021gravitational} and compared our proposed model with this baseline. There were 4 hidden layers with 320 neurons in each. The batch size was $10^{3}$ and the training lasted for $10^{3}$ epochs. For the amplitude network we used the  Adam optimizer \cite{adam} with a learning rate of $10^{-3}$ and the activation function was ReLU \cite{relu}. For the phase network, we used the Adamax \cite{adam} optimizer with a learning rate of  $10^{-2}$ and the activation function was softplus \cite{softplus}. It is worth mentioning that in preliminary experiments various hyperparameters which affect the training process, including learning rate, optimizer type, batch size and learning rate schedule were tested. The best performing set of hyperparameters in terms of MSE on our validation set were finally chosen for the baseline model. As far as preprocessing is concerned, in both cases $\log(q)$ was used as an input instead of $q$, which was then scaled using the  {\tt StandardScaler} from {\tt Scikit-Learn} \cite{sklearn}. At the output, the coefficients were used raw for the amplitude network and were scaled using {\tt Scikit-Learn}'s {\tt MinMaxScaler} for the phase network. All experiments were conducted on an NVIDIA RTX 2080 Ti GPU.

\begin{table}
\caption{\label{tab:MSE} Mean square error (MSE) of the predictions of the ANN surrogate model for the training set and for the validation set, for the baseline network.}
\centering
\begin{tabular}{|c|c|c|}
\hline & MSE for training set & MSE for validation set \\
 & (average of 5 runs) & (average of 5 runs) \\
\hline  Amplitude & $1.79 \times 10^{-7} \pm 3.52 \times 10^{-10}$ & $1.84 \times 10^{-7} \pm 3.29 \times 10^{-10}$ \\
\hline Phase & $1.05 \times 10^{-8} \pm 2.18 \times 10^{-10}$ & $1.06 \times 10^{-8} \pm 2.11 \times 10^{-10}$ \\
\hline
\end{tabular}

\end{table}

The ANN prediction of the EIM coefficients of the training set waveforms will be denoted as  $\hat{\boldsymbol{y}}_i \equiv \{\hat{\alpha}_{k}(\boldsymbol{\lambda}_i)\}_{k=1}^{n}$. During training, the standard mean square error
\begin{equation}
    MSE = \frac{1}{N} \sum_{i=1}^{N} \| \hat{\boldsymbol{y}}_i - \boldsymbol{y}_i \|_2^2
\end{equation}
was measured and minimized, where the $\| \cdot \|_2$ notation represents the Euclidean norm of a vector. Table \ref{tab:MSE}  displays the MSE (average of 5 runs) of the predictions of the ANN surrogate model for the training set and the corresponding predictions for the validation set, for the baseline network. The MSEs are in the range $\sim 10^{-8}-10^{-7}.$ During our experiments it was observed that the amplitude required less epochs to converge (about 400) whereas the phase network always took more (about 800). Throughout our ablation study, we closely monitored the behavior of both the MSE and the mean absolute error (MAE). After careful consideration, we determined that these metrics exhibit similar trends. As a result, we chose to report the MSE in our experiments, as it is the objective function that we utilized to train our ANNs.

In Table \ref{tab:mismatch} we display the calculated mismatch $\mathcal{M}$ of the predictions of the baseline network for the validation set (average of 5 runs) in order to outline how we evaluated the performance of our ANN surrogate model by comparing the mismatch of the generated waveforms with the original waveforms from the SEOBNRv4 model, which served as our benchmark for accuracy assessment. The maximum mismatch is $7.73 \times 10^{-3}$, the $95^{\text {th}}$ percentile is $2.95 \times 10^{-4}$ and the median mismatch is $8.39 \times 10^{-5}$. These values are only several times (less than an order of magnitude) larger than the corresponding mismatch values for the ground truth in Table \ref{mismatch-dist-train}, demonstrating that the ANN surrogate model can predict waveforms for arbitrary $\boldsymbol{\lambda}$ (within the training ranges) with good accuracy\footnote{The accuracy of the SEOBNRv4 model itself is between $10^{-2}$ and $10^{-4}$.}.

\begin{figure}[ht!]
\includegraphics[width=10cm]{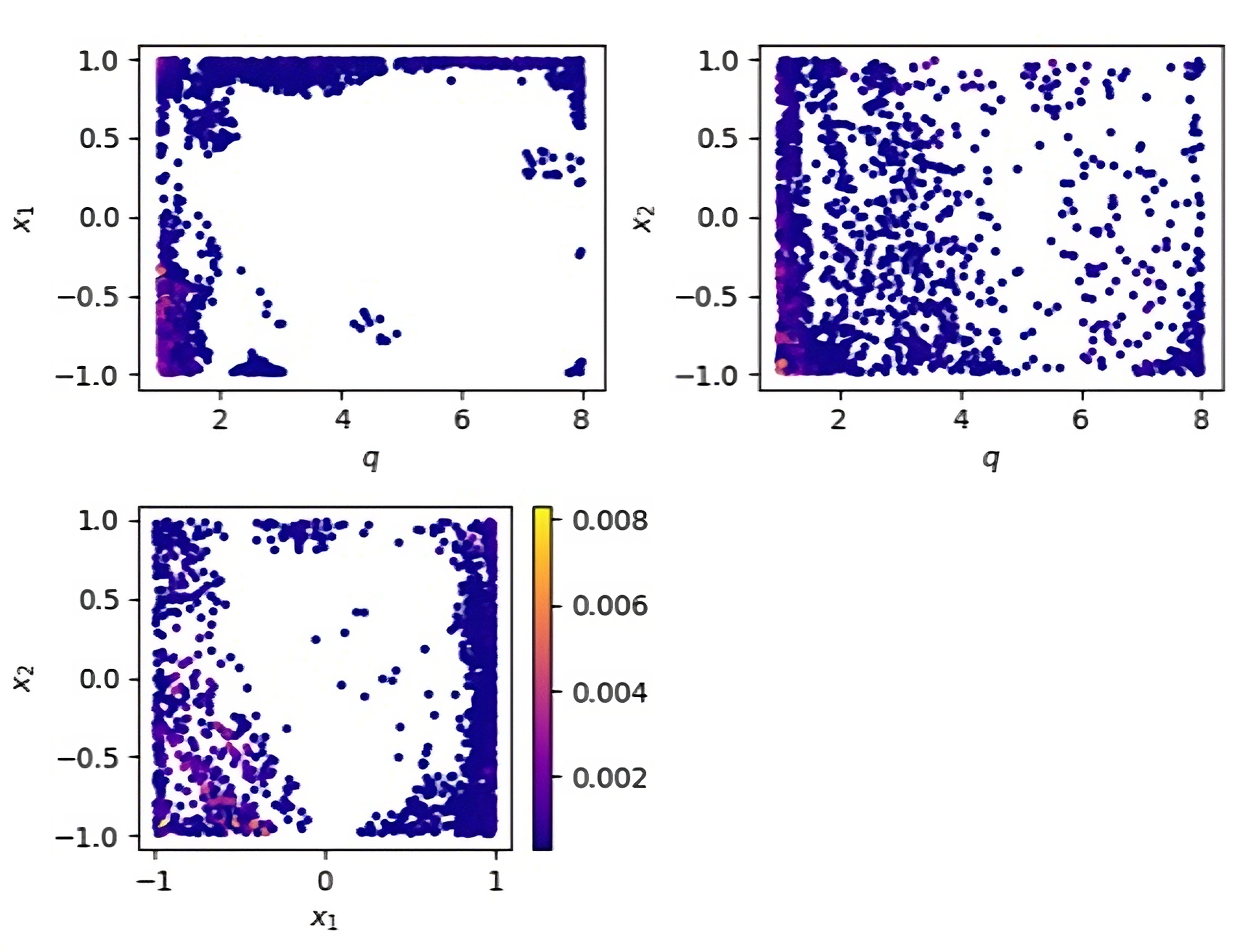}
\centering
\caption{Two-dimensional and three-dimensional distribution of mismatch values larger than $\sim 3\times 10^{-4}$ ($95^{th}$ percentile) of the predictions of the baseline network for the waveforms in the validation set.  }
\label{mismatch-dist-predict}
\end{figure}

The two-dimensional and three-dimensional distributions of mismatch values larger than $\sim 3\times 10^{-4}$ ($95^{th}$ percentile) are shown in Fig. \ref{mismatch-dist-predict}. Similarly to the ground-truth distributions of Fig. \ref{mismatch-dist-train}, large mismatches are seen primarily for high spin values and at the corners and edges of the two-dimensional distributions. 

\begin{table}
\caption{\label{tab:mismatch} Mismatch $\mathcal{M}$ of the predictions of the baseline network for the validation set (average of 5 runs). The minimum, maximum, $95^{\text {th }}$ percentile and median mismatch are shown.
}
\centering
\begin{tabular}{|c|c|}
\hline & Mismatch $\mathcal{M}$ \\
 & (average of 5 runs) \\
\hline Min & $2.70 \times 10^{-6} \pm 3.43 \times 10^{-7}$ \\
Max & $7.73 \times 10^{-3} \pm 5.38 \times 10^{-4}$ \\
$95^{\text {th }}$ & $2.95 \times 10^{-4} \pm 6.61 \times 10^{-6}$ \\
Median & $8.39 \times 10^{-5} \pm 1.91 \times 10^{-6}$ \\
\hline
\end{tabular}

\end{table}

\begin{figure}[ht!]
\includegraphics[width=14cm]{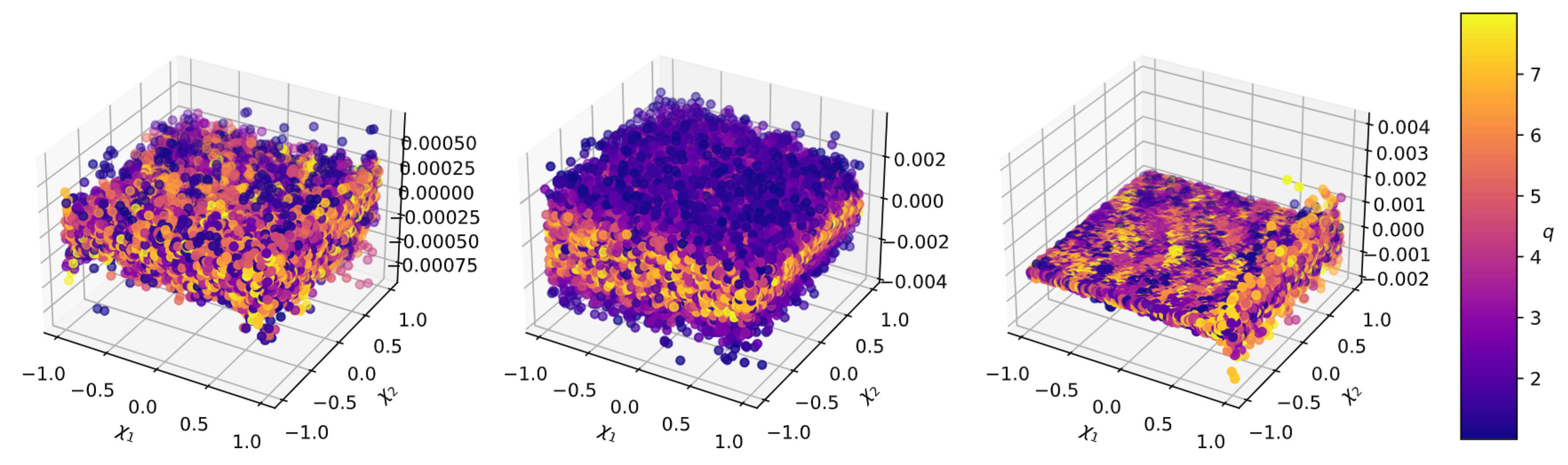}
\centering
\caption{Dependence of the residual error for three selected EIM coefficients for the amplitude on the input parameters $\boldsymbol{\lambda}=\{\chi_1, \chi_2, q\}$ (the dependence on $q$ is shown using a colormap). The example in the middle panel shows a strong dependence on the mass ratio $q$, whereas the example on the right shows a large residual error at the largest value of $\chi_1$.}
\label{amp errors}
\end{figure}

\begin{figure}[ht!]
\includegraphics[width=11cm]{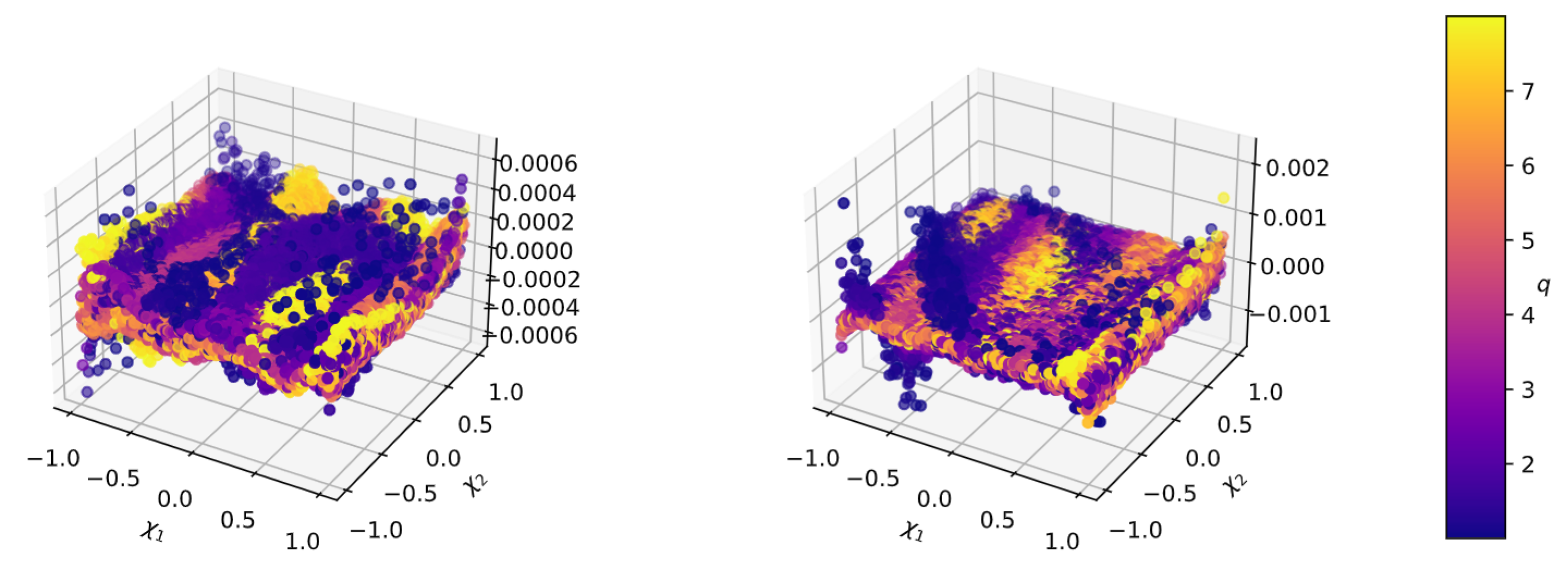}
\centering
\caption{Dependence of the residual error for two selected EIM coefficients for the phase on the input parameters $\boldsymbol{\lambda}=\{\chi_1, \chi_2, q\}$ (the dependence on $q$ is shown using a colormap). Some clustering of the errors as a function of the mass ratio can be seen.}
\label{phi errors}
\end{figure}

\begin{figure}[ht!]
\includegraphics[width=13cm]{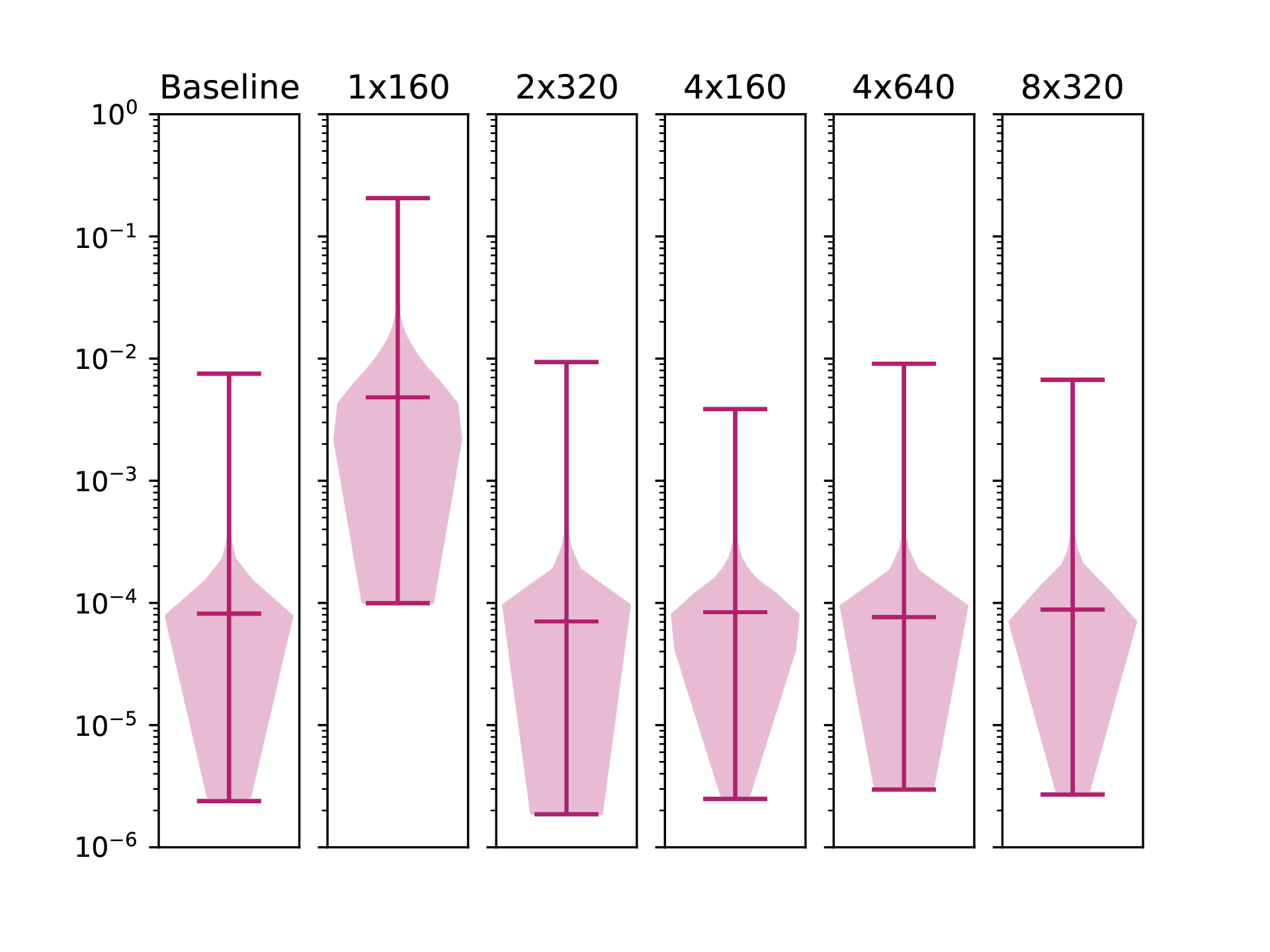}
\centering
\caption{Violin plots comparing the mismatches (for the validation set) between the various baseline network architectures. The middle horizontal line marks the median and the extent of the lines show the minimum and maximum values. In each panel, the envelope is proportional to the density of points. Above the plots the first number corresponds to the number of hidden layers while the second is the number of nodes in each hidden layer.}
\label{baseline-architecture-no-res}
\end{figure}

{ \subsubsection{Baseline model architecture exploration}
Another direction that was explored was that of the architecture of the baseline model. Two types of experiments were tried out, one of shallower models and one of deeper. Both the number of hidden layers and nodes per hidden layer were part of this experimentation. Specifically, for the shallow networks three scenarios were put to test, the input and output layers were kept the same but the hidden layers were altered to a) a single layer with 160 nodes, b) 2 layers with 320 nodes each, c) 4 layers with 160 nodes and for the deeper architecture version hidden layers were altered to a) 4 layers with 640 nodes and b) 8 layers with 320 nodes each.  The final mismatches for these experiments are presented in Table~\ref{tab:mm arch no res} and their corresponding violin plots are shown in Figure~\ref{baseline-architecture-no-res}. We choose violin plots with a logarithmic scale to visually compare the methods, as the errors are very small and cover several orders of magnitude. All of the evaluated architectures achieve more or less the same results, with one exception: the shallow network with one layer of 160 neurons performs significantly worse.}

{
\begin{table} [ht!]
\caption{\label{tab:mm arch no res} Mismatch $\mathcal{M}$ of the predictions of the different architectures of the baseline network for the validation set (average of 5 runs). The maximum, $95^{\text {th }}$ percentile and median mismatch are shown.}
\centering
\resizebox{1 \textwidth}{!}{
\begin{tabular}{|c|c |c| c |c|} 
 \hline
  $\:$ &  \multicolumn{3}{|c|} {Mismatch $\mathcal{M}$ (average of 5 runs)} \\  
   $\:$ &   \multicolumn{1}{|c}{Max}  &  \multicolumn{1}{c}{$95^{\rm th}$ percentile} & Median\\  
 \hline
 {baseline}&   $7.73\times10^{-3} \pm 5.38\times10^{-4}$  & $2.95\times10^{-4} \pm 6.61\times10^{-6}$  & $8.39\times10^{-5} \pm 1.91\times10^{-6}$\\ 
 \hline
  {1x160}&  $2.00\times10^{-1} \pm 7.59\times10^{-3}$  & $2.65\times10^{-2} \pm 9.54\times10^{-4}$  & $4.77\times10^{-3} \pm 1.68\times10^{-4}$  \\ 
 \hline
  {2x320}&  $8.84\times10^{-3} \pm 1.31\times10^{-3}$ & $3.07\times10^{-4} \pm 3.85\times10^{-5}$  & $7.43\times10^{-5} \pm 6.88\times10^{-6}$ \\ 
 \hline 
 {4x160}&  $4.57\times10^{-3} \pm  7.11\times10^{-4}$  & $8.51\times10^{-4} \pm 4.75\times10^{-4}$  & $2.65\times10^{-4} \pm 1.47\times10^{-4}$  \\ 
 \hline
  {4x640}& $8.44\times10^{-3} \pm 2.94\times10^{-3}$  & $3.07\times10^{-4} \pm 2.07\times10^{-5}$ & $8.67\times10^{-5} \pm 2.39\times10^{-6}$ \\
 \hline 
  {8x320}&  $8.39\times10^{-3} \pm 1.59\times10^{-3}$   & $2.89\times10^{-4} \pm 1.71\times10^{-5}$ & $7.69\times10^{-5} \pm 2.64\times10^{-6}$  \\

 \hline
\end{tabular}}

\end{table}
}

\section{Improvement through Residual Errors Network}
\label{sec: alternations}

\begin{figure}[ht!]
\includegraphics[width=10cm]{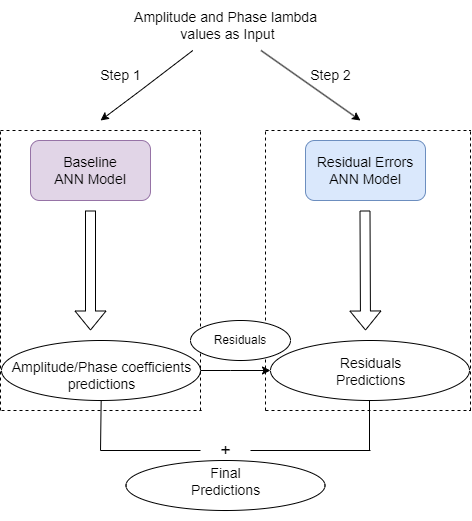}
\centering
\caption{A block diagram of the implemented models.}
\label{Model_schematic}
\end{figure}

Having established the baseline ANN surrogate model in Sec. \ref{sec:baseline_impl}, we proceeded in implementing and evaluating three different pathways for improving the mismatch. First, by using a residual error modeling approach, second, using auxiliary tasks to avoid overfitting, and finally by constructing a feature space that can be better exploited by DL models. The implemented pipeline is outlined in Figure \ref{Model_schematic}. Both models take as input the parameters $\boldsymbol{\lambda}$. The baseline models are trained first, to predict amplitude and phase coefficients. Then, the residual errors are computed as shown in Eq.~\eqref{eq:residual_errors} and the residual model is trained to predict these. The final predictions are the sum of the outputs of the two models.

The first improvement (which also turned out to be the most significant) was obtained by adding a second network (after the training) which can make predictions for the errors of the first network.  We remind that for the baseline model, the first network  had three-dimensional input $\{\boldsymbol{\lambda_i}\}_{i=1}^{N}$  and produced predictions $\hat{\boldsymbol{y}}(\boldsymbol{\lambda})$ (for an arbitrary $\boldsymbol{\lambda}$) with 18 dimensions for the amplitude and 8 for the phase network. For all $\{\boldsymbol{\lambda}_i\}_{i=1}^{N}$ in the training set, one can obtain the corresponding predictions  \{$\hat{\boldsymbol{y}}(\boldsymbol{\lambda}_i)\}_{i=1}^{N}$ and calculate the {\it residual} 
\begin{equation}
    \label{eq:residual_errors}
    \boldsymbol{e}_i \equiv \boldsymbol{y}(\boldsymbol{\lambda}_i) - \hat {\boldsymbol{y}}(\boldsymbol{\lambda}_i),
\end{equation}
where, as already defined, $\boldsymbol{y}$ is the ground truth. Note that the input $\boldsymbol{\lambda}$ undergoes various transformations before being fed to the network, including logarithmic transform of q, and min-max normalization. The second network was designed to have the {\it same input and architecture as the first network}, but this time it was trained on the residuals $\boldsymbol{e}_i$ (which were first scaled using ``MinMaxScaler'' from \texttt{scikit-learn} \cite{sklearn})  to make predictions for the residual $\hat {\boldsymbol{e}}(\boldsymbol{\lambda}) $ at an arbitrary $\boldsymbol{\lambda}$. For this  second network the MSE was defined as
\begin{equation}
    MSE = \frac{1}{N} \sum_{i=1}^{N} \| \hat{\boldsymbol{e}}_i - \boldsymbol{e}_i \|_2^2 .
\end{equation} 
Adding the prediction $\hat {\boldsymbol{e}}$ for the residual to the  prediction $\hat{\boldsymbol{y}}$ of the first network, one obtains an {\it improved prediction}  
\begin{equation}
    \tilde{\boldsymbol{y}} \equiv \hat{ \boldsymbol{y}} +  \hat{\boldsymbol{e}}.
\end{equation}

\begin{table} [ht!]
\caption{\label{tab:amp res} Mean square error (MSE) of the predictions of the ANN surrogate model for the amplitude and phase of the validation set, using the baseline network (middle column) and when a second network that models the residual error is added (right column). In the second case, the MSE of the phase is reduced by a factor of $~\sim 5$.}
\centering
\begin{tabular}{|c |c |c|} 
 \hline
  $\:$ & MSE for baseline network & MSE with residual network\\   & (average of 5 runs) & (average of 5 runs) \\
\hline
 amplitude  & $1.84\times10^{-7} \pm 1.90\times10^{-10}$ &  $1.80\times10^{-7} \pm 2.74\times10^{-10}$  \\ 
 \hline 
phase & $1.08\times10^{-8}  \pm  2.11\times10^{-10}$ &  $1.93\times10^{-9}  \pm  1.58\times10^{-12}$  \\ 
  \hline
\end{tabular}

\end{table}

\begin{table}
\caption{\label{tab:mismatch-res} Mismatch $\mathcal{M}$ of the predictions of the baseline network (middle column) and  when a second network that models the residual error is added (right column), for the validation set (average of 5 runs). The minimum, maximum, $95^{\text {th }}$ percentile and median mismatches are shown.
}
\centering
\begin{tabular}{|c|c|c|}
\hline & Mismatch $\mathcal{M}$  & Mismatch $\mathcal{M}$ \\
 & (average of 5 runs) & (average of 5 runs) \\
 & baseline network  & with residual network \\
\hline Min & $2.70 \times 10^{-6} \pm 3.43 \times 10^{-7}$ & $1.78 \times 10^{-7} \pm 2.34 \times 10^{-8}$ \\
Max & $7.73 \times 10^{-3} \pm 5.38 \times 10^{-4}$ &  $5.75\times10^{-4} \pm 1.14\times10^{-5}$  \\
$95^{\text {th }}$ & $2.95 \times 10^{-4} \pm 6.61 \times 10^{-6}$ & $1.33\times10^{-4} \pm 1.01\times10^{-7}$  \\
Median & $8.39 \times 10^{-5} \pm 1.91 \times 10^{-6}$ & 
$4.33\times10^{-5} \pm 2.73\times10^{-8}$ \\
\hline
\end{tabular}

\end{table}

It is worth noting that the residual errors $\boldsymbol{e}_i$ for the EIM coefficients of the $N$ waveforms in the training set are not always distributed randomly, but can show a certain structure. Fig. \ref{amp errors} displays the dependence of the residual error for three selected EIM coefficients for the amplitude on the input parameters $\boldsymbol{\lambda}=\{\chi_1, \chi_2, q\}$ (the dependence on $q$ is shown using a colormap). The example in the middle panel shows a strong dependence on the mass ratio $q$, whereas the example on the right shows a large residual error at the largest value of $\chi_1$.
The distribution of the residual errors for other coefficients in the EIM expansion is quite similar to one of these characteristic cases. Fig. \ref{phi errors} shows the corresponding dependence of the residual error for two selected EIM coefficients for the phase. Some clustering of the errors as a function of the mass ratio can be seen, which is a strong indication that a second network can learn the residual error.

Table \ref{tab:amp res} displays the mean square error (MSE) of the predictions of the ANN surrogate model for the amplitude and phase of the validation set, using the baseline network (middle column) and when a second network that models the residual error is added (right column). For the amplitude there is only minimal improvement, but for the phase, the addition of the second network that models the residual error reduces the MSE of the predictions for the validation set by a considerable factor of $~\sim 5$.

\begin{figure}[ht!]
\includegraphics[width=7cm]{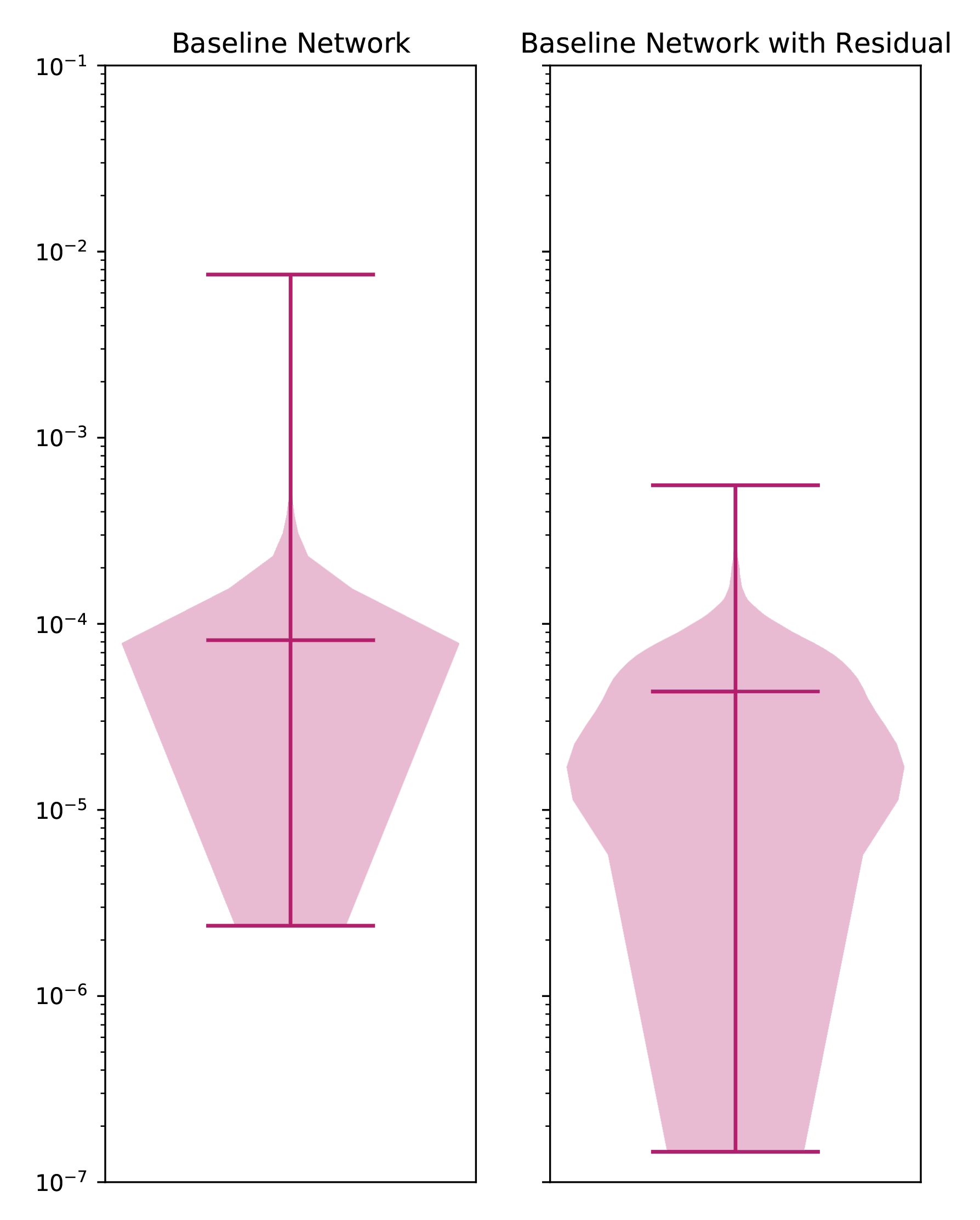}
\centering
\caption{Violin plots comparing the mismatches (for the validation set) between the baseline network and the case when a second network that models the residual error is added. The middle horizontal line marks the median and the extent of the lines show the minimum and maximum values. In each panel, the envelope is proportional to the density of points. A significant reduction of the mismatch is achieved when the network for the residual error is added.}
\label{res-violin}
\end{figure}

In Table \ref{tab:mismatch-res}, we display the mismatch $\mathcal{M}$ of the predictions of the baseline network (middle column) and when a second network that models the residual error is added (right column), for the validation set (average of 5 runs). In the second case, the minimum error is reduced by a factor of $\sim 15$, the maximum error by a factor of $\sim 13$, and the $95^{\text {th }}$ percentile and median by a factor of $\sim 2$.
Fig. \ref{res-violin} compares the mismatches (for the validation set) between the baseline network and the case when a second network that models the residual error is added, as a violin plot. The middle horizontal line marks the median and the extent of the lines show the minimum and maximum values. In each panel, the envelope is proportional to the density of points. 

The significant improvements in the mismatch shown in Table \ref{tab:mismatch-res} and Fig. \ref{res-violin} demonstrate that adding a second network that learns the residual errors is indeed beneficial to the construction of surrogate models for gravitational waves from BBH inspiral. There is good reason to hope that this strategy will also prove to be beneficial for other types of GW template banks, such as binary neutron star inspiral waveforms.

\begin{figure}[ht!]
\includegraphics[width=12cm]{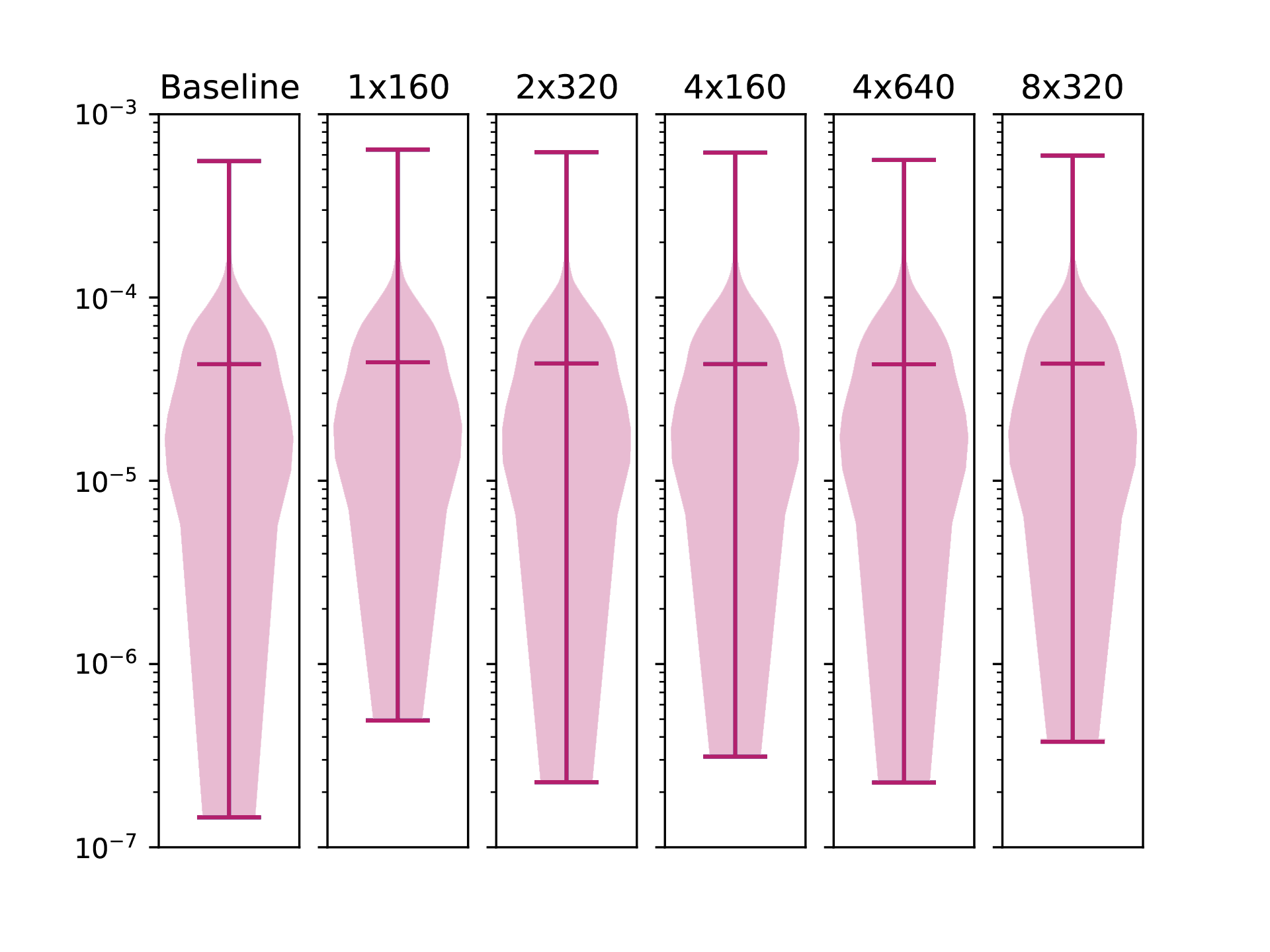}
\centering
\caption{Violin plots comparing the mismatches (for the validation set) between the various baseline network architectures with the effect of the residual errors network. The middle horizontal line marks the median and the extent of the lines show the minimum and maximum values. In each panel, the envelope is proportional to the density of points. Above the plots the first number corresponds to the number of hidden layers while the second is the number of nodes in each hidden layer.}
\label{baseline-architecture-with-res}
\end{figure}

{
\subsection{Effect of residual network on different baseline architecture models}
The idea of residual errors network was also tried during the experiments which concerned the architecture of the baseline model. In all cases the architecture of the residual errors network was not altered and kept with 4 hidden layers with 320 nodes in each one of them. The final mismatches for these experiments with the addition of the residual errors network are presented in Table~\ref{tab:mm arch with res} and their corresponding violin plots are shown in Figure~\ref{baseline-architecture-with-res}. As shown the best choice for baseline model is that of 4 hidden layers with 320 nodes in each of them, followed by a residual errors network with the same architecture.

\begin{table} [ht!]
\caption{\label{tab:mm arch with res} Mismatch $\mathcal{M}$ of the predictions of the different architectures of the baseline network for the validation set (average of 5 runs) with the effect of the residual errors network. The maximum, $95^{\text {th }}$ percentile and median mismatch are shown.}
\centering
\resizebox{1 \textwidth}{!}{
\begin{tabular}{|c|c |c| c |c|} 
 \hline
  $\:$ &  \multicolumn{3}{|c|} {Mismatch $\mathcal{M}$ (average of 5 runs)} \\  
   $\:$ &   \multicolumn{1}{|c}{Max}  &  \multicolumn{1}{c}{$95^{\rm th}$ percentile} & Median\\  
 \hline
 {baseline}&   $5.75\times10^{-4} \pm 1.14\times10^{-5}$  & $1.33\times10^{-4} \pm 1.01\times10^{-7}$  & $4.33\times10^{-5} \pm 2.73\times10^{-8}$\\ 
 \hline
  {1x160}&  $6.65\times10^{-4} \pm 4.72\times10^{-5}$  & $1.38\times10^{-4} \pm 1.71\times10^{-6}$  & $4.46\times10^{-5} \pm 4.48\times10^{-7}$  \\ 
 \hline
  {2x320}&  $5.88\times10^{-4} \pm 1.94\times10^{-5}$ & $1.33\times10^{-4} \pm 3.25\times10^{-7}$  & $4.35\times10^{-5} \pm 1.51\times10^{-7}$ \\ 
 \hline 
 {4x160}&  $5.96\times10^{-4} \pm  2.07\times10^{-5}$  & $1.79\times10^{-4} \pm 9.21\times10^{-5}$  & $7.05\times10^{-5} \pm 5.39\times10^{-5}$  \\ 
 \hline
  {4x640}& $6.12\times10^{-4} \pm 2.85\times10^{-5}$  & $1.34\times10^{-4} \pm 27.87\times10^{-7}$ & $4.37\times10^{-5} \pm 1.81\times10^{-7}$ \\
 \hline 
  {8x320}&  $5.80\times10^{-4} \pm 3.27\times10^{-5}$   & $1.33\times10^{-4} \pm 2.82\times10^{-7}$ & $4.33\times10^{-5} \pm 1.07\times10^{-7}$  \\
 \hline
\end{tabular}}

\end{table}

\subsection{Time measurements}

In this section we measure the time required to generate 1000 coefficients in a single forward pass, for the baseline model ($t_b$) and the entire system including the residual network ($t_r$). The results are shown in Table~\ref{tab:times}, where all measurements were made 200 times and the mean is reported. The measurements are made for the amplitude networks, and similar times should be observed for the phase ones. The CPU used is an Intel(R) Core(TM) i7-9700K CPU @ 3.60GHz, and the GPU used is a NVIDIA RTX 2080 Ti GPU. The number of threads used is 8. The fastest network by far is the shallowest (one hidden layer of 160 neurons), whereas the slowest is the widest (4 hidden layers of 640 neurons), followed by the deepest (8 hidden layers of 320 neurons). The residual network always has the same architecture of 4 hidden layers with 320 neurons each, for fair comparison between experiments. Note that, even though the fastest network remains the fastest after the addition of the residual network, the difference in run times between each model and the baseline $4\times 320$ decreases. Finally note that optimizations, like torchscript\footnote{https://pytorch.org/docs/stable/jit.html} or explicitly threaded network calls, might further decrease the overhead of the residual network.

\begin{table}[]
    \centering
    \caption{Time measurements for various baseline architectures before and after the addition of the residual network. Measurements are made in deployment mode, i.e., no gradients are computed, for a batch size of 1000.}
    \label{tab:times}
    \begin{tabular}{|c|c|c|c|}
        \hline device & model & $t_b$ (ms) & $t_r$ (ms) \\ \toprule
        CPU & $4\times 320$ & 1.66 & 3.75 \\
        & $2\times 320$ & 0.78 & 3.88  \\
        & $4\times 160$ & 0.62 & 3.66  \\
        & $1\times 160$ & 0.24 & 2.04  \\
        & $8 \times 320$  & 3.44 & 6.40 \\
        & $4\times 640$ & 6.59 & 7.70 \\ \midrule
        GPU & $4\times 320$ & 0.27 & 0.54 \\
        & $2\times 320$ & 0.17 & 0.45  \\
        & $4\times 160$ & 0.23 & 0.50  \\
        & $1\times 160$ & 0.11 & 0.37  \\
        & $8 \times 320$ & 0.44 & 0.69 \\
        & $4\times 640$ & 0.61 & 0.72  \\ \bottomrule
    \end{tabular}
    
\end{table}

Based on its performance and time, before and after the addition of the residual network, we choose the $4\times 320$ baseline network to further explore our proposed input and output manipulations, described in the following sections.

}

\section{Exploration of Feature Space and Output Manipulation}
\label{sec: exploration}
\subsection{Feature Space Manipulation}
\label{sec: Feature Space Manipulation}

While working with the baseline network, we aimed for manipulations of the input space of the networks that would lead to a better mismatch. In some cases the input feature space dimension was increased and in some others the total size of the training dataset was increased by inserting an altered branch of samples.

\vspace{0.5cm}

{\it a. Exploitation of similarities between waveforms.} One such idea was to enlarge the input parameter space to four-dimensional, by adding a new parameter that describes physical relations between different waveforms. At 1.5 post-Newtonian (PN) order the time evolution of the frequency $f=\Omega/\pi$ (where $\Omega$ is the orbital frequency) of the quadrupolar part of the waveform is described by \cite{ flanagan, Kidder_etal_1993}:
\begin{equation}
    \frac{df}{dt} = \frac{96}{5} \pi^{8/3} \mathcal{M}_c+c^{5/3} f^{11/3} \left[1 -  \left(  
    \frac{743}{336} + \frac{11 \mu}{4 M} \right) x + (4 \pi - \beta) x^{3/2} + O(x^2)  \right],
    \label{freq_evolution}
\end{equation}
where $M=m_1+m_2$ is the \textit{total} mass of the binary system, $\mu = m_1 m_2 /M$ is the \textit{reduced} mass, $\mathcal{M}_c=\mu^{3/5} M^{2/5}$ is the \textit{chirp} mass and $x = (\pi M f)^{2/3}$. At this order, the effect of the two spins $\chi_1, \chi_2$ on the waveform, enters through a single parameter $\beta$, which is essentially a mass-ratio-weighted sum of the two spins:
\begin{equation}
\beta \equiv \left( \frac{113}{12} + \frac{25}{4q} \right)  \frac{q^2}{(1+q)^2} \chi_1 + \left( \frac{113}{12} + \frac{25q}{4} \right) \frac{1}{(1+q)^2} \chi_2 .
 \label{beta}
\end{equation}
Two different waveforms, characterized by different $\boldsymbol{\lambda}=\{q, \chi_1, \chi_2\}$ values can thus have the same $\beta$ value, leading to identical waveforms at 1.5PN order (and rather similar waveforms, overall).
So, the idea was to insert this new parameter, which has physical meaning, in order to help the network learn the interpolation coefficients. 

\vspace{0.5cm}

{\it b. Augmentation of the training set.} Next, we tried to improve the problem of the presence of the worst mismatches at boundary values of the mass ratio and spins. As a remedy for the large mismatch when $q=1$ was approached, we first tried to augment the dataset with additional input samples $1/q, \chi_2, \chi_1$. corresponding the same coefficients with $q, \chi_1, \chi_2$, which, however, did not yield significant improvement. Next, we tried $\log(q)$ and $-\log(q)$ in place of $1/q$, which gave better results.

\vspace{0.5cm}

{\it c. Dissection of the input space.} 
Another tactic that was followed was that of dissecting the input feature space into a number of $K$ groups and evaluating the performance of the networks in each group. To that end,
the input was divided to $K=2$ groups according to the value of the mass ratio $q$ and two separate networks were trained for both cases of amplitude and phase, each followed by its corresponding residual errors network. The hope was that the individual networks would produce smaller mismatches, by focusing on a smaller  range of $q$, specifically $q \in [1, 4.2]$ for the first group and $q \in [3.8, 8]$ for the second group. Note that the $q$ ranges have a small overlap.

\begin{table} [ht!]
\caption{\label{tab:amp all} Mean square error (MSE) of the predictions of different variants of the ANN surrogate model for the amplitude of the validation set, without (middle column) and with the addition of a network that models the residual error (right column). The smallest mismatch is shown in boldface. See the main text for the explanation of the different variants.}
\centering
\begin{tabular}{|c |c |c|} 
 \hline
   & MSE (average of 5 runs) & MSE (average of 5 runs) \\   &
without residual network & with residual network \\
 \hline
 baseline  & $1.84\times10^{-7} \pm 1.90\times10^{-10}$ &  $1.80\times10^{-7} \pm 2.74\times10^{-10}$  \\ [1ex] 
 \hline 
 
  $\beta$ parameter & $1.82\times10^{-7} \pm 2.44\times10^{-10}$ &  $1.78\times10^{-7} \pm 3.84\times10^{-10}$  \\ [1ex] 
 \hline 
 
  -$\log(q)$  & $1.77\times10^{-7} \pm 4.02\times10^{-10}$ &  $1.77\times10^{-7} \pm 3.17\times10^{-10}$  \\ [1ex] 
 \hline 
 
  K-networks  & $1.86\times10^{-7} \pm 1.25\times10^{-10}$  &   ${\bf 1.74\times10^{-7}} \pm 8.03\times10^{-11}$  \\ [1ex] 
 \hline
  Net per coef  & ${\bf 1.75\times10^{-7}} \pm 2.26\times10^{-10}$ &  $1.75\times10^{-7} \pm 2.38\times10^{-10}$  \\ [1ex] 
 \hline 
 
  $f(1-\boldsymbol{y})$ output& $1.90\times10^{-7} \pm 1.79\times10^{-10}$ &  $1.82\times10^{-7} \pm 5.47\times10^{-10}$  \\ [1ex] 
 \hline
\end{tabular}

\end{table}

\subsection{Output Manipulation}

\vspace{0.5cm}

{\it a. Dedicated network per output coefficient.}
In an effort to achieve smaller mismatches by manipulating the output, we used a dedicated training network for each coefficient. For the baseline case examined in this work, we therefore used 18 networks for the amplitude and 8 networks for the phase. When the residuals were also modeled, an equal number of dedicated networks (one per residual) was also added.

\vspace{0.5cm}

{\it b. Output augmentation.}
Finally, another idea to push the network to learn the desired output, was to insert a new branch with a function $f(\boldsymbol{y})$. For that reason, the quantity $(1-\boldsymbol{y})$ was added
as an extra output for the training network. The final prediction $\hat{\boldsymbol{y}}$ was obtained by combining the predictions for $\boldsymbol{y}$ and $(1-\boldsymbol{y})$.

\vspace{0.5cm}

In all of the above cases, we also experimented with adding a network (or dedicated networks) for modeling the residual error. The results of our experiments are presented in the following Section.

\begin{table} [ht!]
\caption{\label{tab:phi all}As in Table \ref{tab:amp all}, but for the   {\it phase}.}
\centering
\begin{tabular}{|c |c |c|} 
 \hline
   & MSE (average of 5 runs) & MSE (average of 5 runs) \\   &
without residual network & with residual network \\
 \hline
 baseline  & $1.08\times10^{-8}  \pm  2.11\times10^{-10}$ &  $1.93\times10^{-9}  \pm  1.58\times10^{-12}$  \\ [1ex] 
 \hline 
 
  $\beta$ parameter & $9.97\times10^{-9}  \pm  8.59\times10^{-11}$ &  $1.94\times10^{-9}  \pm  1.50\times10^{-12}$  \\ [1ex] 
 \hline 
 
  $-\log(q)$  & $1.04\times10^{-8}  \pm 1.09\times10^{-10}$ &  $2.35\times10^{-9}  \pm  5.89\times10^{-11}$  \\ [1ex] 
 \hline 
 
  $K$-networks  &  ${\bf 4.47\times10^{-9}}  \pm  2.37\times10^{-10}$ &  $2.94\times10^{-9}  \pm 8.51\times10^{-11}$  \\ [1ex] 
 \hline
  Net per coef  & $1.58\times10^{-8}  \pm  4.73\times10^{-10}$ &  $1.96\times10^{-9} \pm  1.72\times10^{-11}$ \\ [1ex] 
 \hline 
 
  $f(1-\boldsymbol{y})$ output& $1.37\times10^{-8}  \pm  3.40\times10^{-10}$ &  ${\bf 1.92\times10^{-9}}  \pm  1.22\times10^{-12}$ \\ [1ex] 
 \hline
\end{tabular}

\end{table}

\begin{table} [ht!]
\caption{\label{tab:mm all} Mismatch $\mathcal{M}$ of the predictions of different variants of the ANN surrogate model (average of 5 runs). For each variant, the maximum, $95^{\text {th }}$ percentile and median mismatches are shown, with the addition of a network that models the residual errors as well as without it. The smallest mismatch is shown in boldface.}
\centering
\resizebox{1 \textwidth}{!}{
\begin{tabular}{|c|c |c| c |c|} 
 \hline
  $\:$ & Residual&  \multicolumn{3}{|c|} {Mismatch $\mathcal{M}$ (average of 5 runs)} \\  
   $\:$ & network&  \multicolumn{1}{|c}{Max}  &  \multicolumn{1}{c}{$95^{\rm th}$ percentile} & Median\\  
 \hline
 {baseline}& No  &  $7.73\times10^{-3} \pm 5.38\times10^{-4}$  & $2.95\times10^{-4} \pm 6.61\times10^{-6}$  & $8.39\times10^{-5} \pm 1.91\times10^{-6}$\\ [1ex] 
                    & Yes & $5.75\times10^{-4} \pm 1.14\times10^{-5}$  & $1.33\times10^{-4} \pm 1.01\times10^{-7}$  & ${\bf 4.33\times10^{-5}} \pm 2.73\times10^{-8}$  \\ [1ex]
 
 \hline
  {$\beta$ parameter}& No  &  $6.25\times10^{-3} \pm 5.17\times10^{-4}$  & $2.68\times10^{-4} \pm 6.72\times10^{-6}$  & $7.80\times10^{-5} \pm 1.26\times10^{-6}$  \\ [1ex] 
                              & Yes &  $5.65\times10^{-3} \pm 8.76\times10^{-6}$ & ${\bf 1.32\times10^{-4}} \pm 2.91\times10^{-7}$  & $4.33\times10^{-5} \pm 7.02\times10^{-8}$ \\ [1ex] 

 \hline
  {$-\log(q)$}& No &  $5.57\times10^{-3} \pm 3.28\times10^{-4}$ & $2.41\times10^{-4} \pm 2.45\times10^{-6}$  & $7.10\times10^{-5} \pm 3.00\times10^{-6}$ \\ [1ex] 
                    & Yes &   ${\bf5.62\times10^{-4}} \pm 6.17\times10^{-5}$  & $1.42\times10^{-4} \pm 3.24\times10^{-6}$  & $4.58\times10^{-5} \pm 4.40\times10^{-7}$  \\ [1ex]
 \hline 
 
 {$K$-nets}& No  &  $2.12\times10^{-3} \pm  1.24\times10^{-4}$  & $1.67\times10^{-4} \pm 2.18\times10^{-6}$  & $5.25\times10^{-5} \pm 8.34\times10^{-7}$  \\ [1ex] 
 
                    & Yes &  $1.12\times10^{-3} \pm 1.15\times10^{-4}$  & $1.49\times10^{-4}   \pm 6.37\times10^{-7}$  & $4.68\times10^{-5} \pm 3.37\times10^{-7}$  \\ [1ex] 
 \hline
  {Net per coef}& No &  $5.50\times10^{-2} \pm 2.33\times10^{-2}$  & $1.43\times10^{-3} \pm 4.63\times10^{-4}$ & $1.89\times10^{-4} \pm 2.55\times10^{-5}$ \\ [1ex] 
                    & Yes &  $7.32\times10^{-4} \pm 1.60\times10^{-4}$   & $1.38\times10^{-4} \pm 5.43\times10^{-6}$  & $4.43\times10^{-5} \pm 6.83\times10^{-7}$  \\ [1ex] 
 \hline 
 
  {$1-\boldsymbol{y}$}& No  &  $6.67\times10^{-3} \pm 1.16\times10^{-3}$   & $2.95\times10^{-4} \pm 7.45\times10^{-6}$ & $8.51\times10^{-5} \pm 1.66\times10^{-6}$  \\ [1ex] 
  
                    & Yes & $5.80\times10^{-4} \pm 1.37\times10^{-5}$  & ${\bf 1.32\times10^{-4}} \pm 1.55\times10^{-7}$  & ${\bf 4.33\times10^{-5}} \pm 5.40\times10^{-8}$  \\ [1ex]
 \hline
\end{tabular}}

\end{table}

\section{Performance of different variants of the ANN surrogate model}
\label{sec: results}

Table \ref{tab:amp all} shows the MSE of the predictions of the different variants of the ANN surrogate model discussed in Sec. \ref{sec: exploration} for the amplitude of the validation set. Without a residual network, the variants with the $\beta$ parameter, the $1/q$ and $-\log(q)$ input augmentations and the individual networks per coefficient show a slight improvement in the mismatch. The variants with the dissection into $K=2$ networks and the output augmentation show a slight worsening, instead. Adding a network that models the residual error slightly improves or does not affect the MSE, except for the case of the  $1/q$  input augmentation, where the MSE becomes slightly worse. Out of all variants shown in Table \ref{tab:amp all}, the dissection into $K=2$ networks with residual error modeling shows the smallest MSE, which is $\sim 5\%$ ($\sim 3\%$) smaller than the MSE of the baseline network without (with) residual modeling. It is important to note that our method is compatible and thus can be used on top of any ANN architecture.  Moreover, our experimental study has demonstrated notable enhancements when compared to the baseline model proposed by \cite{khan2021gravitational} which served as the starting point of our experimentation. 

\begin{figure}
     \centering
     \begin{subfigure}[ht!]{0.8\textwidth}
         \centering
         \includegraphics[width=\textwidth]{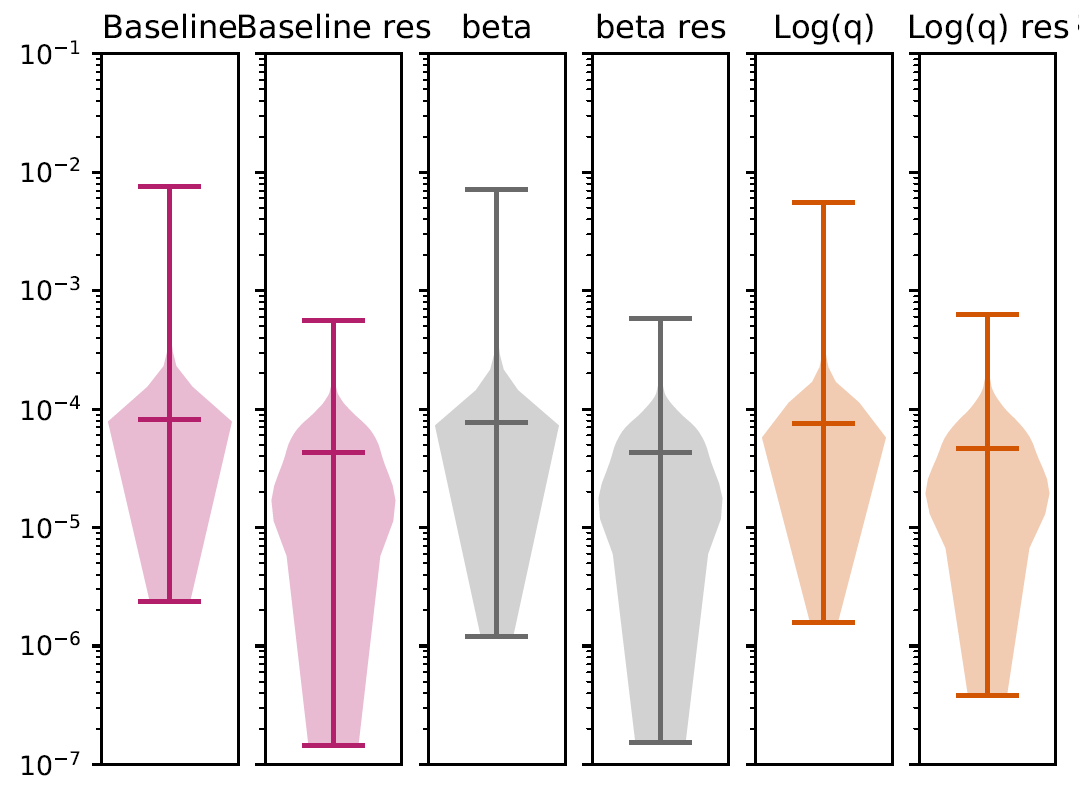}
     \end{subfigure}
     \hfill
     \begin{subfigure}[ht!]{0.8\textwidth}
         \centering
         \includegraphics[width=\textwidth]{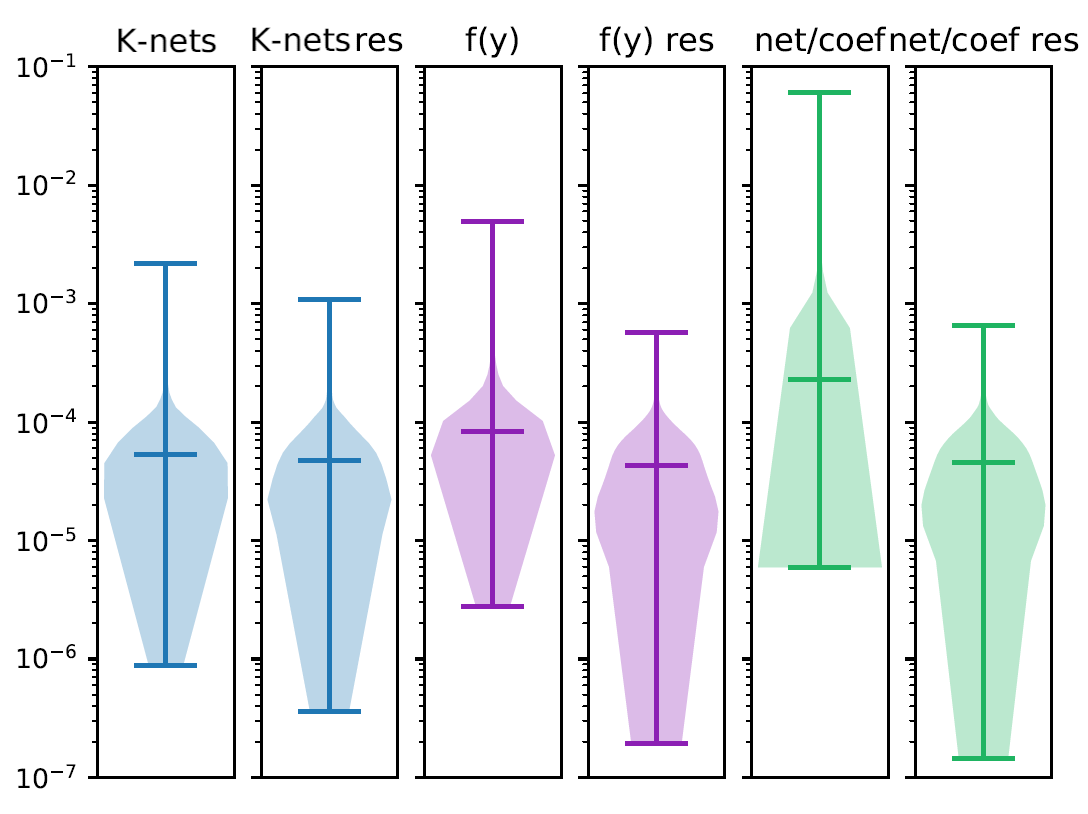}
     \end{subfigure}
     \hfill
        \caption{Violin plots comparing the mismatches for different variants  of the ANN surrogate model without (left panel for each variant) and with the addition of a network that models the residual error (right panel for each variant).The middle horizontal line marks the median and the extent of the lines show the minimum and maximum values. In each panel, the envelope is proportional to the density of points. A significant reduction of the mismatch is achieved in several variants.  See text for details.}
        \label{violin couples}
\end{figure}

Table \ref{tab:phi all} shows the corresponding MSE for the phase. Without a residual network, the variants with the $q$ input, the
$\beta$ parameter, the $-\log(q)$ input augmentation and the dissection into $K=2$ networks show improvements between $\sim 4\%$ and $\sim 60\%$, whereas the remaining variants show a worsening of the MSE.  When a network that models the residual error is added, the MSE is reduced significantly for all variants. The highest reduction, by a factor of $\sim 7$, can be observed for the output augmentation. This is the best case out of all variants shown in Table \ref{tab:phi all}, with the baseline network with added residual error modeling coming a close second.

In Table \ref{tab:mm all} we show the mismatch $\mathcal{M}$ of the predictions of the different variants of the ANN surrogate model discussed in Sec. \ref{sec: exploration}  (the average of 5 runs is shown). For each variant, the maximum, $95^{\text {th }}$ percentile and median mismatches are shown, both with and without the addition of a network that models the residual errors. The corresponding violin plots are shown in Fig. \ref{violin couples} (except for the case of $q$ input). For each variant (shown in different colors) the left (right) panel displays the case without (with) a network for the residual error. The middle horizontal line marks the median and the extent of the lines show the minimum and maximum values. The envelope is proportional to the density of points. 

From the mismatch results displayed in Table  \ref{tab:mm all} and in Fig. \ref{violin couples}, it is evident that the most significant improvements to the ANN surrogate model come from the inclusion of the network that models the residual error. Both the minimum and maximum error are reduced by more than an order of magnitude, compared to the baseline network without residual error modeling. Specifically, the maximum error is reduced by a factor of $\sim 17$ for the variant with $q$ input and residual error modeling, compared to the baseline network. The minimum error is reduced by a factor of $\sim 15$ for the baseline with residual error modeling and by a similar factor for the dedicated networks per coefficient, compared to the baseline network without residual error modeling. The $95^{\text {th }}$ percentile mismatch is reduced by a factor of $\sim 2$ with respect to the baseline network without residual error modeling. The smallest $95^{\text {th }}$ percentile mismatches were achieved in the cases of the $\beta$ parameter and of the output augmentation, with the baseline model and the variant with $q$ input coming close second (all cases with residual error modeling added). The best performance of the median mismatch was observed for the baseline network with residual error modeling and for the variants with $q$ input and output augmentation (also with residual error modeling). Compared to the baseline network without residual error modeling, these variants had a median error reduced by a factor of $\sim 2$.

\section{Discussion \& Conclusions}
\label{conclusions}

Deep learning methods have been employed in gravitational-wave astronomy to accelerate the construction of surrogate waveforms for the inspiral of spin-aligned black hole binaries, among other applications. As we demonstrated here, the residual error of an ANN that models the coefficients of the surrogate waveform expansion has sufficient structure to be learnable by a second network. This is especially true for the residual error of the surrogate model coefficients for the phase of the waveform. We added a second ANN (of the same architecture as the main network) and showed that the maximum mismatch for waveforms in a validation set was reduced by more than an order of magnitude. 

Furthermore, we explored several other ideas for improving the accuracy of the surrogate model, such as the exploitation of similarities between waveforms, the augmentation of the training set, the dissection of the input space, using dedicated networks per output coefficient and output augmentation. In several cases, small improvements were observed, but the most significant improvement still came from the addition of a second network that models the residual error. Since the residual error for more general surrogate waveform models (when e.g. eccentricity of tidal effects are included) may also have a specific structure, one can expect our method to be applicable in such more general cases. The gain in accuracy may then lead to significant gains in computational time. We plan to investigate such cases in the future.

Specifically, in a series of extensive experiments, we showed that the proposed bag-of-tricks methods can improve the maximum mismatched between real and reconstructed waveforms by 1.2-1.4 times, using the $-\text{log}(q)$ or $\beta$ parameter method, and up to 3.6 times, using the K-nets method, compared to the mismatch achieved by using the baseline model at \emph{zero} additional computational overhead during deployment. Furthermore, the proposed residual error modeling method can achieve up to 13.4 times improved mismatch at about two times the computational overhead compared to the baseline method. This result is especially noteworthy when taking into consideration the fact that using networks with larger capacity as the baseline coefficient predictors only increases the computational cost without providing proportionate - or any - improvement in terms of mismatch. This means that in order to reach mismatch values of over an order of magnitude lower than the baseline network, regardless of its capacity, a second model that learns the first's residual errors is essential.

While our results demonstrate the effectiveness of the proposed method in accurately predicting waveforms, we also acknowledge certain limitations that could be addressed in future research. Specifically, our study required a substantial amount of training data and the number of parameters was relatively low-dimensional compared to the 7-dimensional space of the general case of merging black holes. To further enhance the proposed approach, we recommend future studies to explore residual error networks and utilize the proposed `bag-of-tricks' method to improve network performance in cases where dense training sets are not available. 

One can also hope to perhaps model the residual errors even further, e.g. by adding a third network to model the ``residual of the residual'', if a clustering persists at the next level. This suggestion is in line with residual connection based architectures \cite{he2016deep}, where each such connection can be viewed as residual error learning from its input to its desired output, i.e., the error of the previous neural structure. 

\section{Acknowledgments}
We are grateful to Theocharis Apostolatos for suggesting to use  the relation for $\beta$ and Leila Haegel, Maria Haney, George Pappas, Vasilis Skliris and Vijay Varma for useful discussions and comments. The authors gratefully acknowledge the Italian Instituto Nazionale di Fisica Nucleare (INFN), the French Centre National de la Recherche Scientifique (CNRS) and the Netherlands Organization for Scientific Research, for the construction and operation of the Virgo detector and the creation and support of the EGO consortium and the COST network CA17137 “G2Net” for support. Panagiotis Iosif acknowledges support by the European Research Council (ERC) under the European Union's Horizon 2020 research and innovation programme under grant agreement No. 759253.

\section{Appendix}
\label{sec:appendix}

\begin{figure}[htb]
    \centering 
\begin{subfigure}[ht]{0.5\textwidth}
  \includegraphics[width=\linewidth]{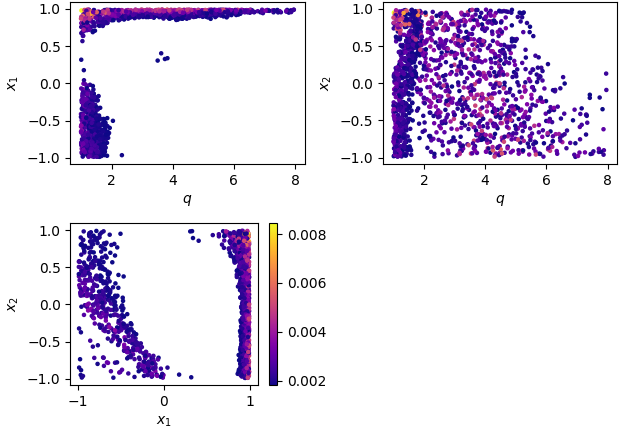}
  \caption{tolerance $10^{-06}$}
  \label{fig:1}
\end{subfigure}\hfil
\medskip
\begin{subfigure}[ht]{0.5\textwidth}
  \includegraphics[width=\linewidth]{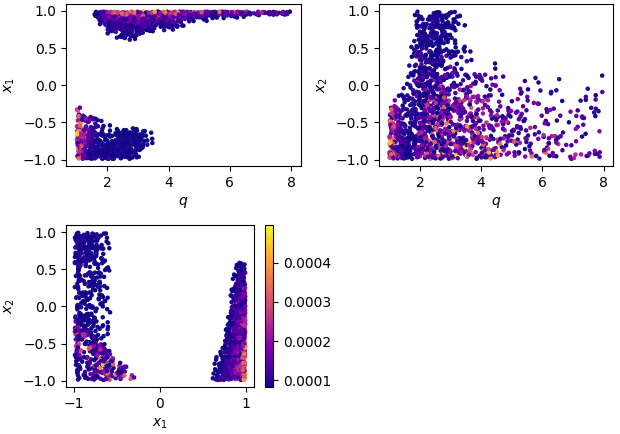}
  \caption{tolerance $10^{-10}$}
  \label{fig:3}
\end{subfigure}

\medskip

\begin{subfigure}[ht]{0.5\textwidth}
  \includegraphics[width=\linewidth]{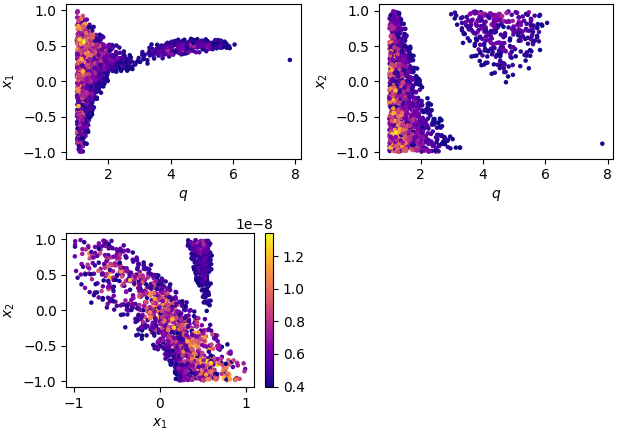}
  \caption{tolerance $10^{-14}$}
  \label{fig:5}
\end{subfigure}\hfil
\medskip
\begin{subfigure}[ht]{0.5\textwidth}
  \includegraphics[width=\linewidth]{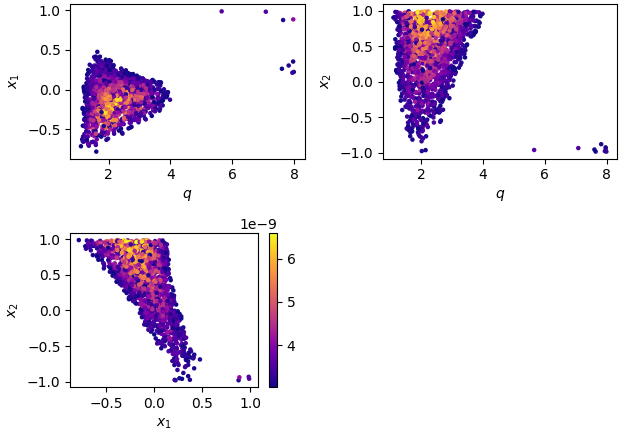}
  \caption{tolerance $10^{-16}$}
  \label{fig:6}
\end{subfigure}
\caption{Two-dimensional and three-dimensional distribution of mismatch values larger than $\sim 8\times 10^{-5}$ ($95^{th}$ percentile) for the waveforms in the validartion set, when reconstructed using the EIM reduced basis via Eq. (\ref{eq:EIM-reconstruction}). The different panels correspond to different values of the greedy tolerance.}
\label{validation_gt_mm}
\end{figure}

 The different panels in Fig.  \ref{validation_gt_mm} display the two-dimensional distributions of mismatch values larger than the $95^{th}$ percentile (for different values of the greedy tolerance) for the waveforms in the validation set, when reconstructed using the EIM reduced basis via Eq. (\ref{eq:EIM-reconstruction}).
 The distributions are different at very small values of the greedy tolerance, when compared to the distributions for higher values of the greedy tolerance.

\newpage

 \bibliographystyle{elsarticle-num} 
\bibliography{main.bib}

\end{document}